\documentclass[a4paper,12pt]{article}
\usepackage[dvips]{color}
\usepackage{ulem}
\usepackage{multirow}
\usepackage{latexsym}
\usepackage{epsfig,epsf,subfig,graphicx,amsfonts,amssymb,color, subfig, color}
\usepackage{hyperref}
\usepackage[english]{babel}
\usepackage[utf8]{inputenc}
\date{}

\title{
Quantitative approach to multifractality induced \\by correlations and broad distribution of data}
\author{Rafa{\l} Rak$^{1,2}$, Dariusz Grech$^3$}

\begin{document}
\maketitle
\noindent
\begin{center}
$^1$\textit{Faculty of Mathematics and Natural Sciences, University of Rzesz\'ow, Pigonia 1,\\ 35-310 Rzesz\'ow, Poland}\\
$^2$\textit{Institute of Nuclear Physics, Polish Academy of Sciences, Radzikowskiego 152,\\ 31-342 Krak\'ow, Poland}\\
$^3$\textit{Institute of Theoretical Physics, University of Wrocław, Pl. M.Borna 9, \\ 50-204 Wrocław, Poland}\\

\end{center}

\begin{abstract}

We analyze quantitatively the effect of spurious multifractality induced by the presence of fat-tailed symmetric and asymmetric probability distributions of fluctuations in time series. In the presented approach different kinds of symmetric and asymmetric broad probability distributions of synthetic data are examined starting from Levy regime up to those with finite variance.
We use nonextensive Tsallis statistics to construct all considered data in order to have good analytical description of frequencies of fluctuations in the whole range of their magnitude and simultaneously the full control over exponent of power-law decay for tails of probability distribution. The semi-analytical compact formulas are then provided to express the level of spurious multifractality generated by the presence of fat tails in terms of Tsallis parameter $\tilde{q}$ and the scaling exponent $\beta$ of the asymptotic decay of cumulated probability density function (CDF).  The results are presented in Hurst and H\"{o}lder languages – more often used in study of multifractal phenomena.  According to the provided semi-analytical relations, it is argued how one can make a clear quantitative distinction for any real data between true multifractality caused by the presence of nonlinear correlations, spurious multifractality generated by fat-tailed shape of distributions - eventually with their asymmetry, and the correction due to linear autocorrelations in analyzed time series of finite length. In particular, the spurious multifractal effect of fat tails is found basic for proper quantitative estimation of all spurious multifractal effects. Examples from stock market data are presented to support these findings.

\end{abstract}
$$
$$
\textbf{Keywords}: multifractality, spurious multifractality, time series analysis, autocorrelations, symmetric and asymmetric distributions, multifractal detrended analysis, generalized Hurst exponent\\
\textbf{PACS:} 05.45.Tp, 89.75.Da, 05.40.-a, 89.75.-k, 89.65.Gh\\
{\it corresponding authors; e-mail: darusz.grech@uwr.edu.pl}

\section{Introduction and motivation}

Multifractality \cite{mf12,mf13,mf14,kantelhardt-preprint,mf1,mf2,mf3} is commonly considered as very interesting feature of complex and composite systems which attracts a lot of attention in many areas of science. The multifractal
properties of time series are extensively studied because of their omnipresence in various phenomena
in nature connected with complexity like turbulence \cite{18,19}, astronomy \cite{20}, climate phenomena \cite{21,22,22a},
physiology \cite{23}, text structure \cite{24,24a,24a1,24b}, physics \cite{25,26} or finances \cite{27,28,29,30,31,32,33,34,35,36,37,38,ft22,ft33,ft44}. This fragmentary list is far from being exhaustive and does not cover enormous number of publications on the subject.

The practical fruits of multifractality are not
precisely known yet but in some fields including finance
interesting features of this phenomenon were shown (see,
e.g.,\cite{mf4,mf5,mf6,mf7,mf8,mf9,mf10,mf11,oswiecimka,czarnecki}) that rise hope for interesting future applications connected with risk analysis.
Therefore the questions regarding accuracy, applicability and reliability of multifractal
measurements are crucial for proper analysis and interpretation of obtained results.

Since
the seminal paper by Kantelhardt, et.al., \cite{Kantelhardt-316} we know that multifractal properties one observes
may appear not only as result of existing long-range nonlinear autocorrelations
but also from the presence of fat tails in probability distributions
of data or from linear autocorrelations present in shorter (finite) time series. The latter effect called also a finite size effect (FSE) has been extensively studied in quantitative way by various authors (see, e.g.,\cite{37,58,60,61,63,64}). In fact the mutual interaction and interplay between these three sources of multifractal effects leads to observable multifractal spectrum. It is a nontrivial task to  determine generally how these three ingredients relatively influence the measured multifractal features. We need some general method and analytical or semi-analytical formulas which could reveal the mutual interplay between true multiscaling of data generated by effects of nonlinear autocorrelations and the remaining sources of multifractality producing in fact multifractal artifacts sometimes called spurious multifractality.
%%%%%%%%%%%%%%%%%%%%%%%%%%%%%%%%%%%%%%%%
The effect of spurious MF can also be a result of the presence of additive white or color noise, short-term memory or periodicity in
multifractal signal \cite{64a,58,64b}. They may significantly change its observed multifractal properties for all data lengths.
%%%%%%%%%%%%%%%%%%%%%%%%%%%%%%%%%%%%%%
Only the multifractality generated by nonlinear effects is most interesting from the practical point of view because it reflects in some sense the genetic structure of complicated intrinsic couplings and information flow inside the complex system. This multiscaling manifests differently at various time scales and actually makes the essence of true multifractal phenomenon.

The
expected level of multifractal artifacts existing due
to finite-size effects and linear autocorrelations in time series was described generally from the quantitative point of view in series of papers \cite{63,64,64a,64b}.
In this paper we will make the similar quantitative analysis of spurious multifractality caused by different types of broad probability distribution of data including also asymmetric fat-tailed distributions. The latter ones are expected to occur in some real systems including financial ones \cite{65,66,67,68}.
It is worth to notice that in fact
nonlinear effects also produce broad distribution of data what in turn influences multifractal phenomena in a way of specific feedback. Hence the
statement that "true" multifractality is generated only by nonlinear
effects is somehow misleading. Nevertheless,  we try to identify in this paper how large the part of multifractality generated by broad distribution of data is -- even if such effects are completely separated from the
nonlinear
correlations, i.e., if the latter are taken to be null, while the
broad distribution is still assumed to exist. This way an additional
contribution to observable multifractal spectrum is produced and one has to be able to determine quantitatively how large this effect is. Hence, in our
approach we will divide the multifractal spectrum into three separate parts having in mind
however, that at least two of them are in fact strongly coupled in any complex system.

We will use the multifractal detrended fluctuation analysis
(MFDFA) \cite{kantelhardt-preprint, Kantelhardt-316} within this paper.  MFDFA is now commonly accepted technique in searching for
multifractal properties of data in time series. MFDFA has been applied so far in diversified scientific problems like, e.g., seismology \cite{seism1,seism2}, cosmology \cite{cosmo}, biology \cite{biol1,biol2}, meteorology \cite{meteo}, medicine \cite{med1,med2}, music \cite{mus1,mus2}, geophysics \cite{geo}, and mainly finances \cite{mf12,32,mf4,mf5,mf6,mf7,mf8,mf9,mf10,mf11}.
 This technique is reported to have an advantage over the other known approach based on wavelets \cite{69}. Since it is described elsewhere (see e.g., \cite{kantelhardt-preprint, oswiecimka,czarnecki,Kantelhardt-316}) we will only briefly recall it here.

 The main steps of MFDFA go as follows. Let $X(j) = \sum_{i=1}^j{(x(i)-<x>)}, \ j = 1,...,M$ be a signal profile, where ${x(i)}_{i=1,...,M}$ is an analyzed time series and $<x>$ denotes averaging over all $i$'s. One divides $X(j)$ into $K_s$ non-overlapping segments of length $s$ (time windows) starting from both the beginning and the end of the signal.
For each segment a local trend $P_{\nu}^{l}$ ($l$-th order polynomial) is estimated and subtracted from the signal profile. Next, for the detrended signal a local variance
$F^2(\nu,s)$ in each segment $\nu$ and $q$-th order fluctuation function  is calculated according to:
\begin{equation}
F_q(s) = \bigg\{ \frac{1}{2 K_s} \sum_{\nu=1}^{2 K_s} [F^2(\nu,s)]^{q/2} \bigg\}^{1/q},
\label{ffun}
\end{equation}
where $q\in\Re$. In this paper, to make the results more readable, we use $l=2$, and apply the scaling range: $s_{min}=40$, $s_{max}=M/20$ (for synthetic data), $s_{max}=M/10$ (for real data). For a signal with fractal characteristics the fluctuation functions exhibit power law scaling:

\begin{equation}
F_q(s) \sim s^{h(q)}
\label{scal}
\end{equation}
where $h(q)$ is a generalized Hurst exponent. The bi- or multifractal stationary signals have $h(q)$ profile as a decreasing function of $q$; if $h(q)=const$ the signal is called monofractal.

Often the multifractal properties are presented in the H\"{o}lder
language as the multifractal singularity spectrum $f(\alpha)$ \cite{mf3}.
The singularity spectrum $f(\alpha)$ can be calculated according to the following relations~\cite{legendre1,legendre2}:
\begin{equation}
\tau(q)=qh(q)-1,~~~~\alpha=\frac{d}{dq}\tau(q)~~\textrm{and}~~f(\alpha)=q\alpha-\tau(q),
\label{scal2}
\end{equation}
where $\alpha$ is called the singularity (H\"{o}lder) exponent.
The wealth of multifractality present in time series can be defined as a spread of the generalized Hurst exponent $\Delta h(q)$. It is considered as dependent on the $q$ parameter range \cite{Kantelhardt-316}:

\begin{equation}
\Delta h(q)=h(q_{min})-h(q_{max})
\label{scal3}
\end{equation}
where $q_{min}$ and $q_{max}$ are respectively the minimal and the maximal value of the real deformation parameter $q$ taken into account (usually the symmetric range $q_{min}=-q_{max}<0$ is proposed).

The degree of multifractality can be also estimated by
measuring the width of $f(\alpha)$ spectrum ~\cite{Kantelhardt-316}:

\begin{equation}
\Delta \alpha=\alpha_{max}(q_{min})-\alpha_{min}(q_{max}).
\label{scal4}
\end{equation}
In the limit $q_{max}=-q_{min}\rightarrow \infty$ both multifractal characteristics in Eqs.(4) and (5) coincide.

Note, that in order to distinguish the deformation multifractal parameter $q$ from the parameter used in "$q$-deformed" fat tailed Tsallis distribution, which will also be used in this paper, the latter one will be denoted further on by $\tilde{q}$.

The paper is organized as follows. In section 2 we investigate the
quantitative effect of broad distribution on multifractal characteristics of data within
MFDFA. Two cases of fat tailed distributions are considered: stable ones
with infinite variance (the Levy type of PDF or CDF) with the attractor made by Levy distribution according
to Generalized Central Limit Theorem  and the
second case -- unstable distributions with finite variance. The are
modeled in this paper by Tsallis $\tilde{q}$-normal distributions ($\tilde{q}$Gaussians). Their attractor, according to
Generalized Central Limit Theorem, is made by Gaussian or Levy distribution and depend on $\tilde{q}$ value. We discuss an impact made on spurious multifractality by symmetric and asymmetric distributions
in both such regimes (Gaussian and Levy). The quantitative findings from synthetic
data obeying these two types of fat tailed CDF are then applied in
section 3 to isolate three types of multifractality (generated respectively by
nonlinear effects, linear effects with FSE and broad distribution of data) in real
financial returns collected for diversified time-lags. Concluding remarks on
the mutual interplay between spurious and real multifractal effects expected
to occur in real time series are summarized in the last section.

\section{Influence of symmetric and asymmetric broad distributions of synthetic data on registered multifractal outcomes}

 To analyze various features of multifractality we will use time series of uncorrelated data drawn from  $\tilde{q}$Gaussian distribution \cite{ts1} as well as time series of empirical, usually nonlinearly correlated data. For all data, we shall explore quantitatively the impact of heavy tailed asymmetric and symmetric probability distributions on multifractality and compare it with the effect of linear and nonlinear correlations present in a signal of finite length (FSE).

The fat tails discovered in the real probability distributions in many complex systems including stock and money market \cite{mf13,ft22,ft33,ft44,ts11} and the complex character of  the  underlying  temporal  correlations indicate that the  conventional concept of ergodicity  may  break down in the real dynamics. Under such conditions the generalized formalism of nonextensive statistical mechanics~\cite{ts1,ts2,ts3,ts4,ts5,ts6} may offer an appropriate framework to generate the corresponding time series.
In non-extensive approach one is capable to pass in a compact and very economic way through all intermediate cases of fat tailed distributions just by altering the value of one  parameter $\tilde{q}$ as described below.
For this reason we use time series sampled from the $\tilde{q}$Gaussian distribution with $\tilde{q}$ being a positive real deformation parameter:
\begin{equation}
p (x) \sim e_{\tilde{q}}^{-a_{\tilde{q}}x^2}=1/[1+({\tilde{q}}-1)a_{\tilde{q}}x^2]^{1/({\tilde{q}}-1)}
\label{qgaussians}
\end{equation}
The cumulative form of $\tilde{q}$Gaussian distribution is defined as follows~\cite{r1}:

\begin{equation}
P_{\pm}(x) =\mathcal{N}_{\tilde{q}}\left(\frac{\sqrt{\pi } ~\Gamma
\left(\frac{1}{2} (3-{\tilde{q}}) ~\beta \right)}{2 ~\Gamma (\beta )~
\sqrt{\frac{\mathcal{B}_{\tilde{q}}}{\beta }}}\pm(x-\bar{\mu}_{\tilde{q}}) \,
   _2F_1(\alpha ,\beta ;\gamma ;\delta )\right)~\label{Pcx}
\end{equation}
where, the $+$ and $-$ signs correspond to the right and left wings of the distribution, while\\
$$
\mathcal{N}_{\tilde{q}}=\left\{
\begin{array}{ccc}
\frac{\Gamma \left[ \frac{5-3{\tilde{q}}}{2-2{\tilde{q}}}\right] }{\Gamma \left[
\frac{2-{\tilde{q}}}{1-{\tilde{q}}}
\right] }\sqrt{\frac{1-{\tilde{q}}}{\pi }\mathcal{B}_{\tilde{q}}} & ~~for  & \tilde{q}<1 \\[3mm]
\frac{\Gamma \left(\frac{1}{{\tilde{q}}-1}\right)}{\Gamma \left(\frac{3-{\tilde{q}}}{2
({\tilde{q}}-1)}\right) \sqrt{\frac{\pi}{({\tilde{q}}-1) \mathcal{B}_{\tilde{q}}}}} & ~~for &
1<{\tilde{q}}<3
\end{array}
\right. ,$$
$$ \bar{\mu} _{\tilde{q}}= \,\int x\frac{\ \left[ p\left( x\right)
\right] ^{\tilde{q}}}{\int \left[ p\left( x\right) \right] ^{\tilde{q}}dx}\
dx\equiv \left\langle x\right\rangle _{\tilde{q}} ,
$$
$$
\mathcal{B}_{\tilde{q}}=\left[ \left( 3-1\right)
\,\bar{\sigma}_{\tilde{q}}^{2}\right] ^{-1},
$$
$\alpha=\frac{1}{2}$, $\beta =\frac{1}{{\tilde{q}}-1}$, $\gamma
=\frac{3}{2}$, $\delta =-\mathcal{B}_{\tilde{q}}({\tilde{q}}-1)(\bar{\mu}_{\tilde{q}}-x)^2$
and $_2F_1(\alpha ,\beta ;\gamma ;\delta)$ is the Gauss hypergeometric function.
This type of family distributions  develops asymptotically (for large $x$) a power law behavior, contrary to exponential behavior characteristic for normal (Gaussian) distribution. Thus, for cumulative $\tilde{q}$Gaussian distributions the relationship
\begin{equation}
P(x)\sim x^{\frac{2}{1-{\tilde{q}}}+1}\sim x^{-\beta}
\label{power}
\end{equation}
holds.
\begin{figure}[h!]
\begin{center}
\vspace*{.3in}
\hspace*{-.1in}
\includegraphics[width=0.9\textwidth, height=0.32 \textwidth]{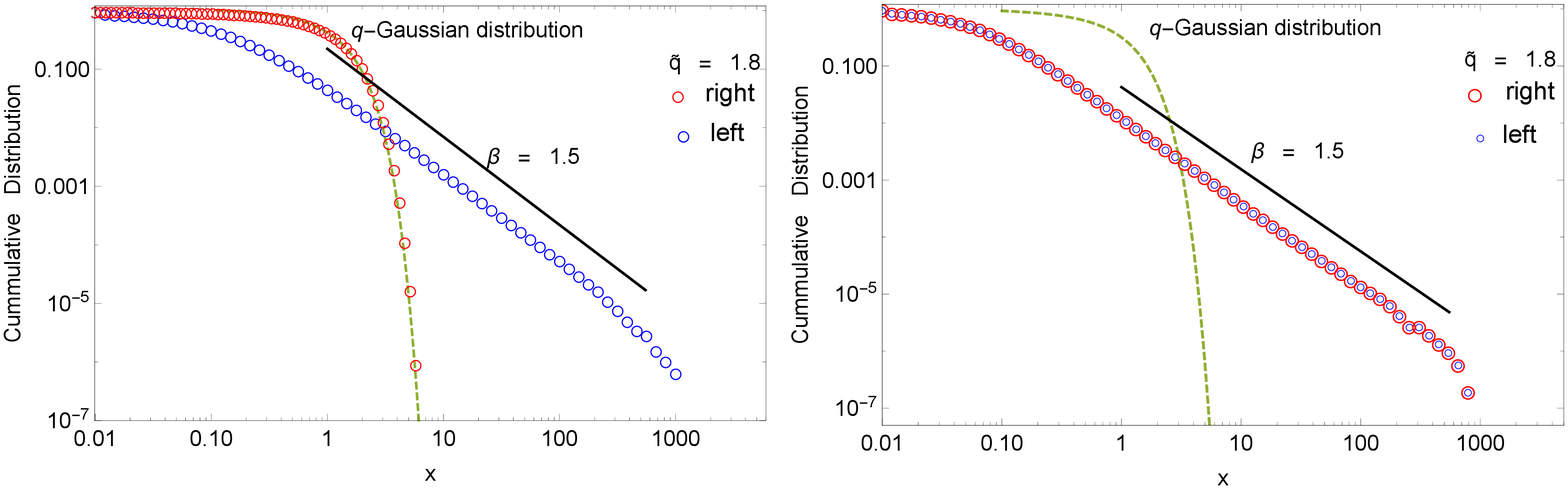}
\includegraphics[width=0.9\textwidth, height=0.32 \textwidth]{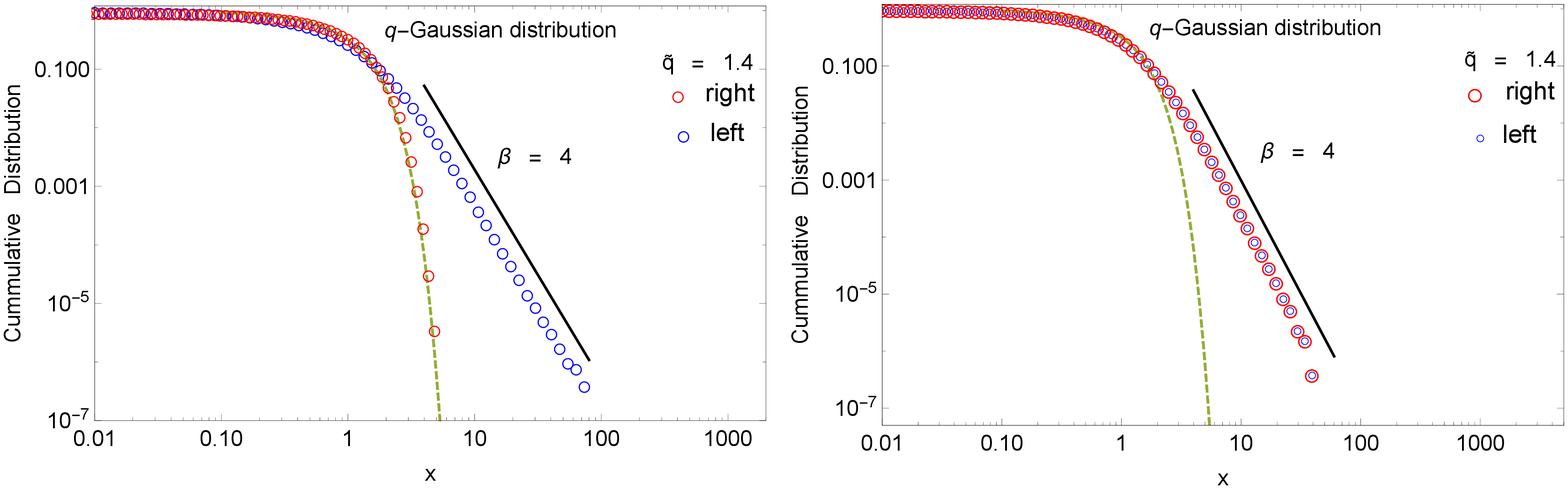}
\end{center}
\caption{Asymmetric (left column) and symmetric (right column) $\tilde{q}$Gaussian cumulative probability distributions for ${\tilde{q}}=1.8$,  (Levy regime) and ${\tilde{q}}=1.4$,  (Gaussian regime) with power law decay of distribution tails $P(x>>1)\sim x^{-1.5}$ and $P(x>>1)\sim x^{-4}$ respectively. The dashed line corresponds to a normal distribution. }
\label{fig1}
\end{figure}
Importantly, distributions of the uncorrelated $\tilde{q}$Gaussian signals, depending on $\tilde{q}$, are either in the Gaussian attractor regime (for $1\leq {\tilde{q}}<5/3$) or Levy attractor regime (for $5/3\leq {\tilde{q}} \leq2$). It means that sum of independent $n$ random variables satisfying such distributions undergoes respectively normal or Levy stable distribution for $n\rightarrow\infty$. In the latter case exponent $\beta$ of the power law in Eq.(8) becomes also the stability parameter of Levy distribution.

\begin{figure}[h!]
\begin{center}
\vspace*{.3in}
\hspace*{-.1in}
\includegraphics[width=1\textwidth, height=0.6 \textwidth]{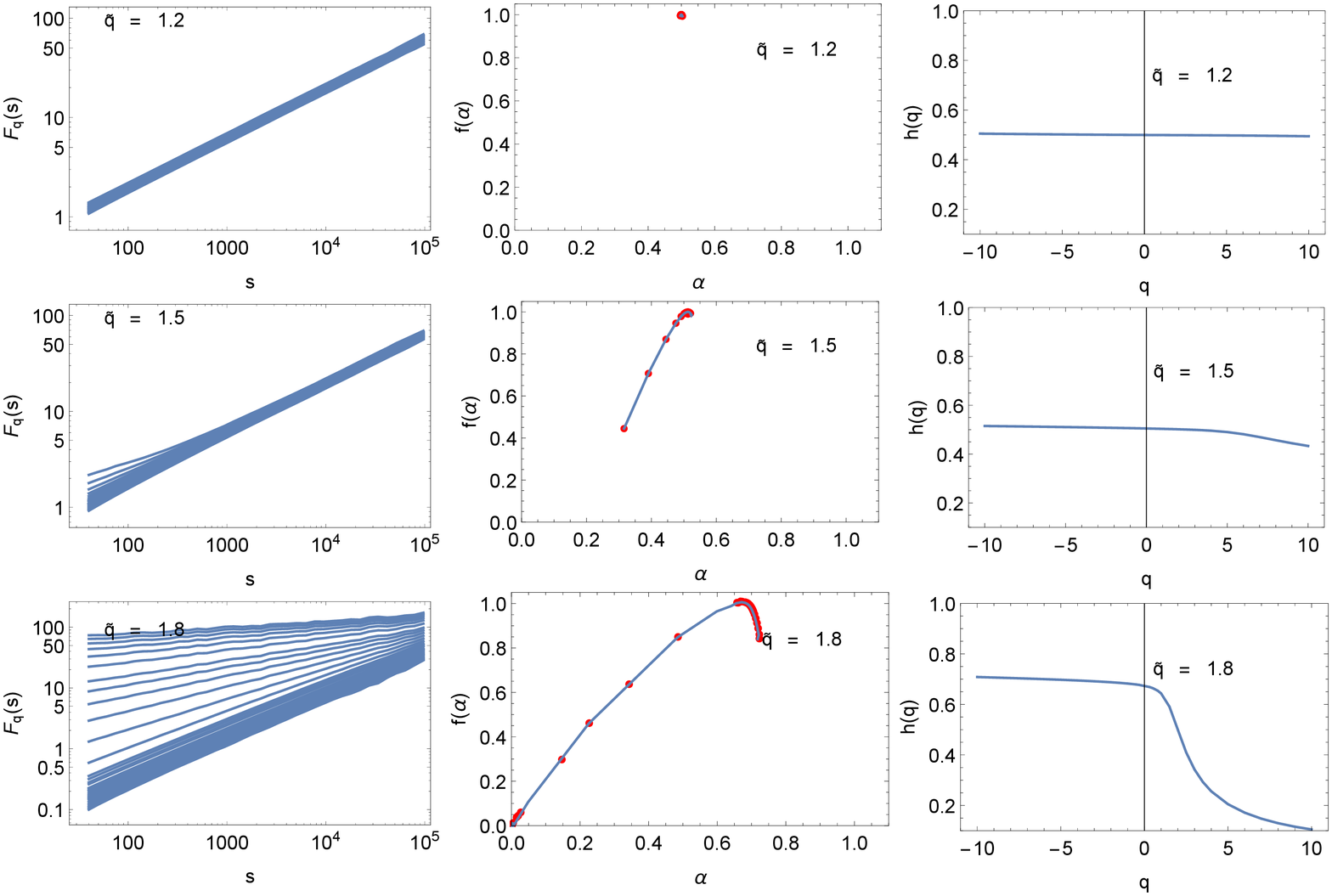}
\end{center}
\caption{Fluctuation function $F_q(s)$ (left column), the multifractal spectrum $f(\alpha)$ (middle column) and the generalized Hurst exponent $h(q)$ (right column) calculated within MFDFA for signals of length $M=10^{10}$ sampled for different $\tilde{q}$ parameters of symmetric $\tilde{q}$Gaussians. In each panel different moments indexed by the integer $-10\leq q\leq 10$ are shown.} \label{fig2}
\end{figure}

\begin{figure}[h!]
\begin{center}
\vspace*{.3in}
\hspace*{-.1in}
\includegraphics[width=1\textwidth, height=0.6 \textwidth]{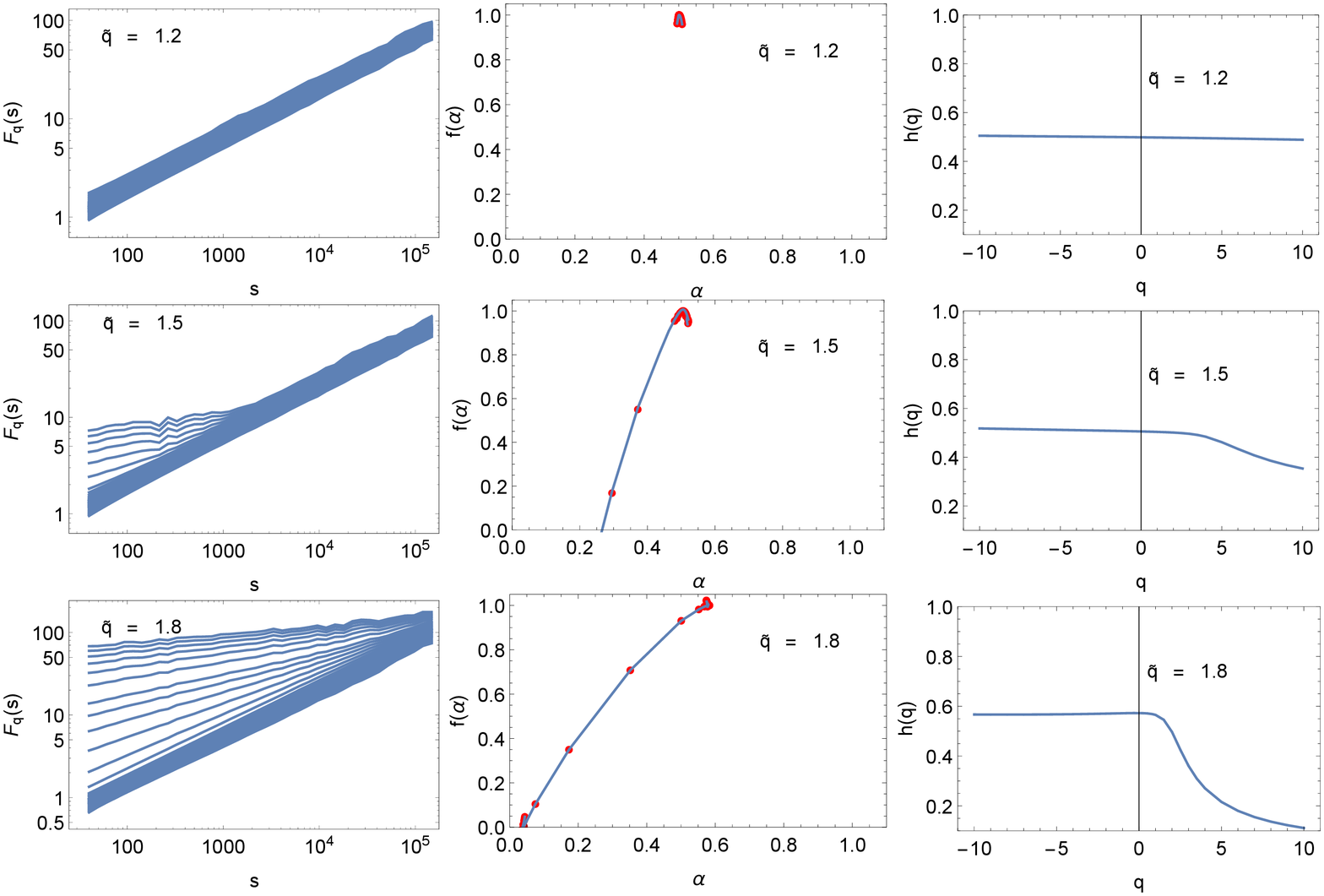}
\end{center}
\caption{Same as in Fig.~\ref{fig2} for asymmetric $\tilde{q}$Gaussians.}
\label{fig3}
\end{figure}

In order to investigate the possible impact of symmetric (asymmetric) broad distributions on multifractal effects, we generated time series with symmetric and asymmetric distribution according to Eq.(\ref{qgaussians}). It was assumed that the asymmetric distribution is one for which the right tail (positive fluctuations) has a normal distribution. This assumption has no impact on final results because MFDFA is invariant under mirror transformation $X(i)\rightarrow -X(i)$. Fig.\ref{fig1} shows an example of symmetric and asymmetric distribution in the Levy (for $\tilde{q}=1.8$) and Gauss (for $\tilde{q}=1.4$) attractor regime respectively.
To avoid the influence of FSE on the registered multifractality of the series, we generated relatively long series, i.e., $M\sim 10^{10}$. According to refs. \cite{63,64} the spurious multifractal spread $\Delta h_{FSE}$ due to presence of FSE is limited then to $\Delta h_{FSE}\lesssim 0.02$ for $q\thickapprox 15$ and $\Delta h_{FSE}\lesssim 10^{-3}$ for $q\thickapprox 2$.
Examples of the fluctuation function ($F_q(s)$), the multifractal spectrum ($f(\alpha)$) and the generalized Hurst exponent ($h(q)$) (both for $\tilde{q}$  from the Gaussian and Levy regimes) for symmetric and asymmetric distributions are shown in Figs.~\ref{fig2} and \ref{fig3}.

For all $\tilde{q}$  a good power-law dependence of $F_q(s)$ was observed for the scales $s\geq1000$. A small variability of the slope coefficients, both for symmetric and asymmetric case in Gaussian regime indicate a monofractal character of the analyzed data in this region. Examples for $\tilde{q}=1.2$ and $\tilde{q}=1.5$  are given for this case in two top rows of Figs.~\ref{fig2}, \ref{fig3}.
Nevertheless the clear crossover $s_X \sim 10^3\div 10^4$ is observed with spurious multifractal effects contributing to multifractal spectrum below $s_X$. A transition from distributions in Gaussian regime to Levy regime attractor corresponding to $\tilde{q}=5/3$ changes also drastically the shape and the spread of the spurious multifractality caused by broad distribution of data (see the bottom row in Figs.\ref{fig2}, \ref{fig3}).
For $\tilde{q}=1.8$ (Levy regime), both for symmetric and asymmetric case,  one observes a good scaling for all scales and for all  values of $q$. Here, however, $F_q$ lines are not parallel especially for large fluctuations --  one clearly observes different slopes for different values of $q$ and this in turn can be wrongly interpreted as the multifractal nature of the analyzed data. In a case of $\tilde{q}$Gaussian distributions from normal attractor region (top two panels of Figs.~\ref{fig2} and \ref{fig3}) the spurious multifractality is mainly generated below the crossover scale $s_X$. Surprisingly this region corresponds not to large but to relatively small fluctuations which contribute to $F_q(s)$ function the most since they are statistically more meaningful in small time windows after detrending procedure. The effect is clearly magnified for asymmetric distributions. The crossover scale $s_X$ slightly moves toward higher values if $\tilde{q}$ increases but details of such relationship are out of the scope of this paper. The comparative results in H\"{o}lder ($f(\alpha)$) and Hurst ($h(q)$) descriptions obtained for $s\geq s_X$ are shown in the second and third columns in Figs.~\ref{fig2} and \ref{fig3}). Distinctly different results for symmetric and asymmetric distributions are observed especially for broad distributions from Levy attractor. It is seen at the level of $f(\alpha)$ for $q=1.8$ where the value of $\alpha_{max}$, i.e., the righthanded edge of spurious multifractal spectrum  is clearly shifted to the left in the case of asymmetrical distribution (compare middle-bottom panels in Figs.~\ref{fig2} and \ref{fig3}).
\begin{footnotesize}
\begin{figure}[h!]
\begin{center}
\vspace*{.3in}
\hspace*{-.1in}
\includegraphics[width=0.8\textwidth, height=0.5 \textwidth]{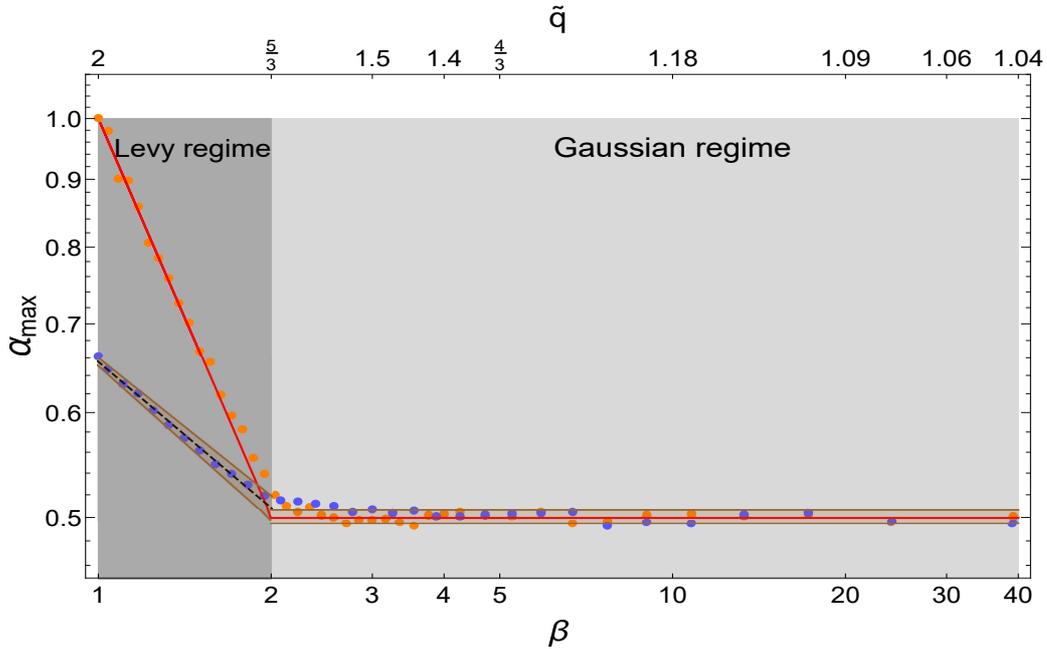}
\end{center}
\caption{
Righthanded edge of spurious multifractality $\alpha_{max}$ generated by the effect of broad PDF. The results are shown in log-linear scale as a function of the $\tilde{q}$Gaussian parameter $\tilde{q}$ and $\beta$ and are calculated for the signals of length $M=10^{10}$ for symmetric (orange symbols) and asymmetric distributions (blue symbols). The red line corresponds to the theoretical value for symmetric distributions ($\alpha_{max}\sim \beta^{-1}$ in the Levy regime; $\alpha_{max}\sim 0.5$ in the Gaussian regime). The black dashed line describes the empirical results for asymmetric distributions, i.e., $\alpha_{max}\sim \beta^{-0.37\pm 0.02}$ in the Levy attractor regime and $\alpha_{max}\sim 0.5$ in Gaussian attractor regime. The shade of bronze around the blue symbols indicates the range of fit accuracy at $CL=95\%$ confidence level. The dark-gray and light-gray area represent the Levy and the Gaussian regime respectively.}
\label{fig4}
\end{figure}
\end{footnotesize}

While in the Gaussian regime the effect of (a)symmetry of probability distribution does not significantly affect multifractal properties ($\alpha_{max}\sim 0.5$), the remarkable differences are visible in the Levy area. For a symmetric case, we observe the dependence $\alpha_{max}(\beta)$ in excellent agreement with theoretical prediction, i.e., $\alpha_{max}\sim \beta^{-1}$~\cite{69}. In contrast, when one of the tails of the distribution has a shape of normal PDF (the asymmetric case) $\alpha_{max}\sim \beta^{-\gamma}$, where ${\gamma=0.37\pm0.02}$ (see Fig.~\ref{fig4}).
The smaller value of this exponent in the latter case indicates that asymmetrical distribution significantly depletes the multifractal nature of the analyzed data. When $\beta$ decreases (the tail of the distribution becomes thicker) the shift of the multifractal spectrum $f(\alpha)$ to the right is slower than in symmetric case.
Looking at $\alpha_{max}$ and its drift along the shape of fat tailed PDF (in here described by $\tilde{q}$ and $\beta$ scaling exponent) one may estimate the influence of fluctuations of particular form on the presence of spurious multifractality generated by the deviation of data from the normal distribution. This effect does not replace however the need to search for full spurious multifractal spectrum generated by the broad form of PDF. It is done below.

To look more closely at the impact of the phenomenon of symmetry, asymmetry and fat tails of probability distributions on measured multifractal features, we study the generalized Hurst exponent dependent on $q$ parameter ($h(q)$) and the properties of its spread $\Delta h(q)$.
\newpage
\begin{figure}[!ht]
\begin{center}
\begin{footnotesize}~~~~Gaussian regime (Symmetric case)~~~~~~~~~~~~~~~~~~~~~~~~~~Gaussian regime (Asymmetric case)\end{footnotesize}\\
\includegraphics[width=0.5\textwidth, height=0.18 \textwidth]{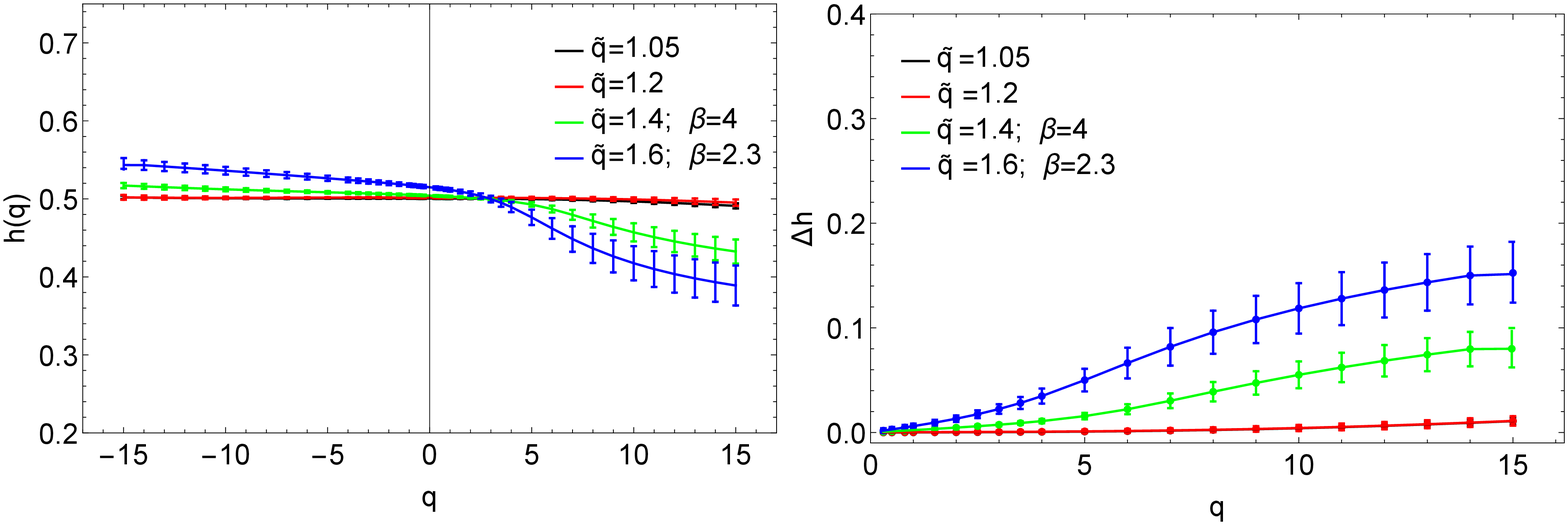}\includegraphics[width=0.5\textwidth, height=0.18 \textwidth]{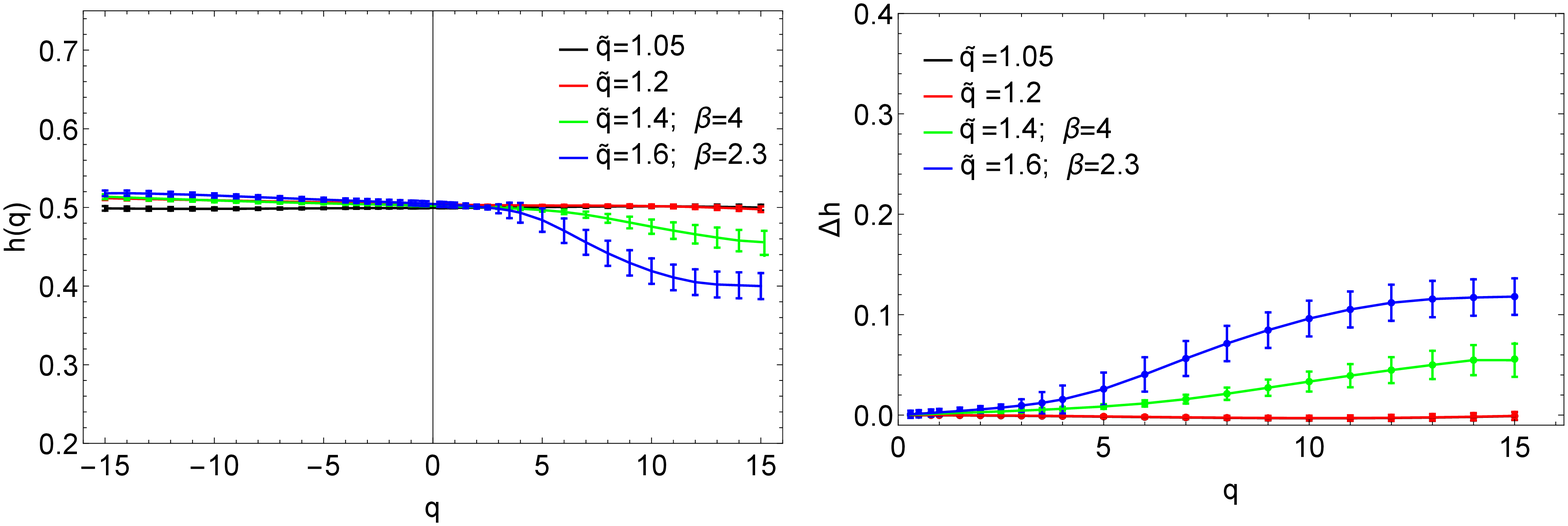}\\
\begin{footnotesize}Levy regime (Symmetric case)~~~~~~~~~~~~~~~~~~~~~~~~~~~~~~~~~Levy regime (Asymmetric case)\end{footnotesize}\\
\includegraphics[width=0.5\textwidth, height=0.18 \textwidth]{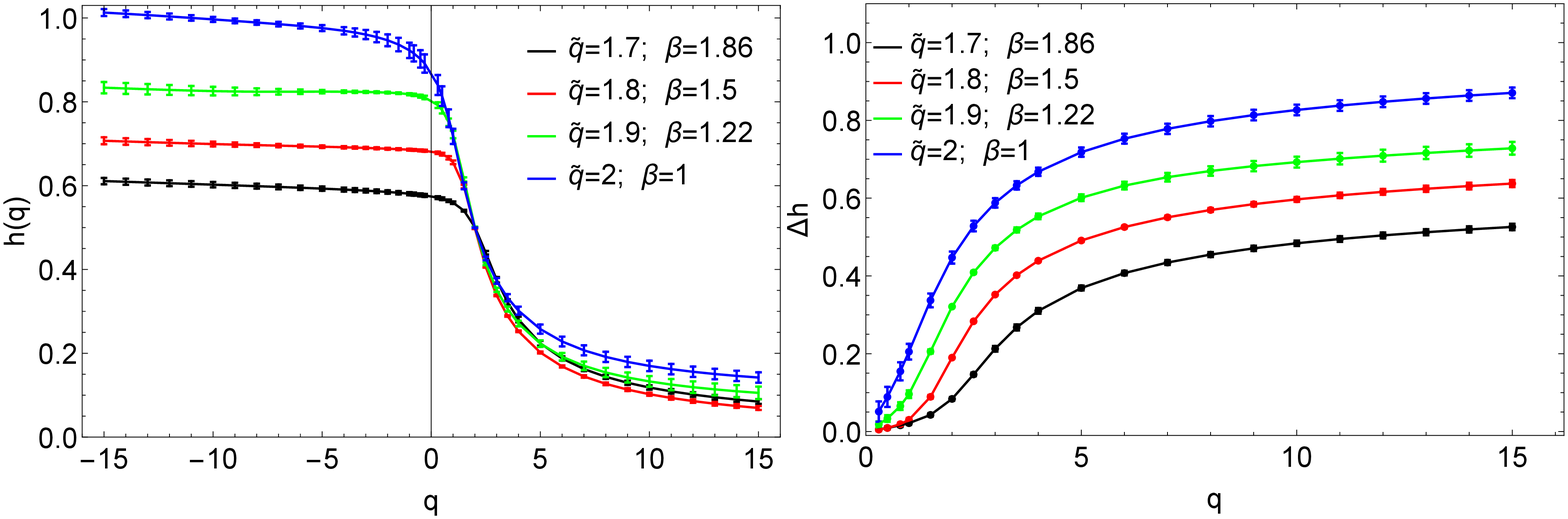}\includegraphics[width=0.5\textwidth, height=0.18 \textwidth]{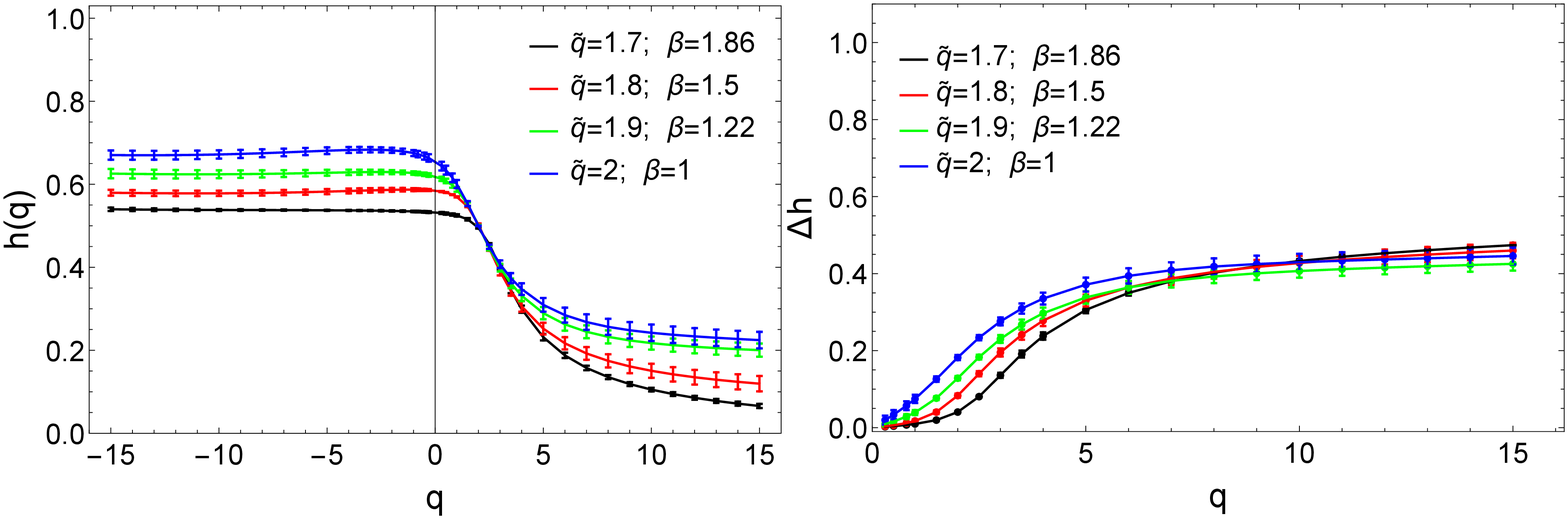}
\end{center}
\begin{footnotesize}
\caption{Generalized Hurst exponent $h(q)$ and its spread $\Delta h$ induced by heavy tails of PDF calculated for time series of length $M=10^{10}$ with data from various symmetric and asymmetric $\tilde{q}$Gaussians.  Examples of both regimes, i.e., the Gaussian ($\tilde{q}=1.05, 1.2, 1.4, 1.6$) attractor regime and the Levy attractor regime ($\tilde{q}=1.7, 1.8, 1.9, 2$) are penetrated as labeled on individual panels. The error bars indicate the range of fit accuracy at
$CL = 95\%$ confidence level.}
\label{fig55}
\end{footnotesize}

\end{figure}

\begin{figure}[!ht]
\begin{center}
\begin{footnotesize}~~~~Gaussian regime (Symmetric case)~~~~~~~~~~~~~~~~~~~~~~~~~~Gaussian regime (Asymmetric case)\end{footnotesize}\\
\includegraphics[width=0.4\textwidth, height=0.27 \textwidth]{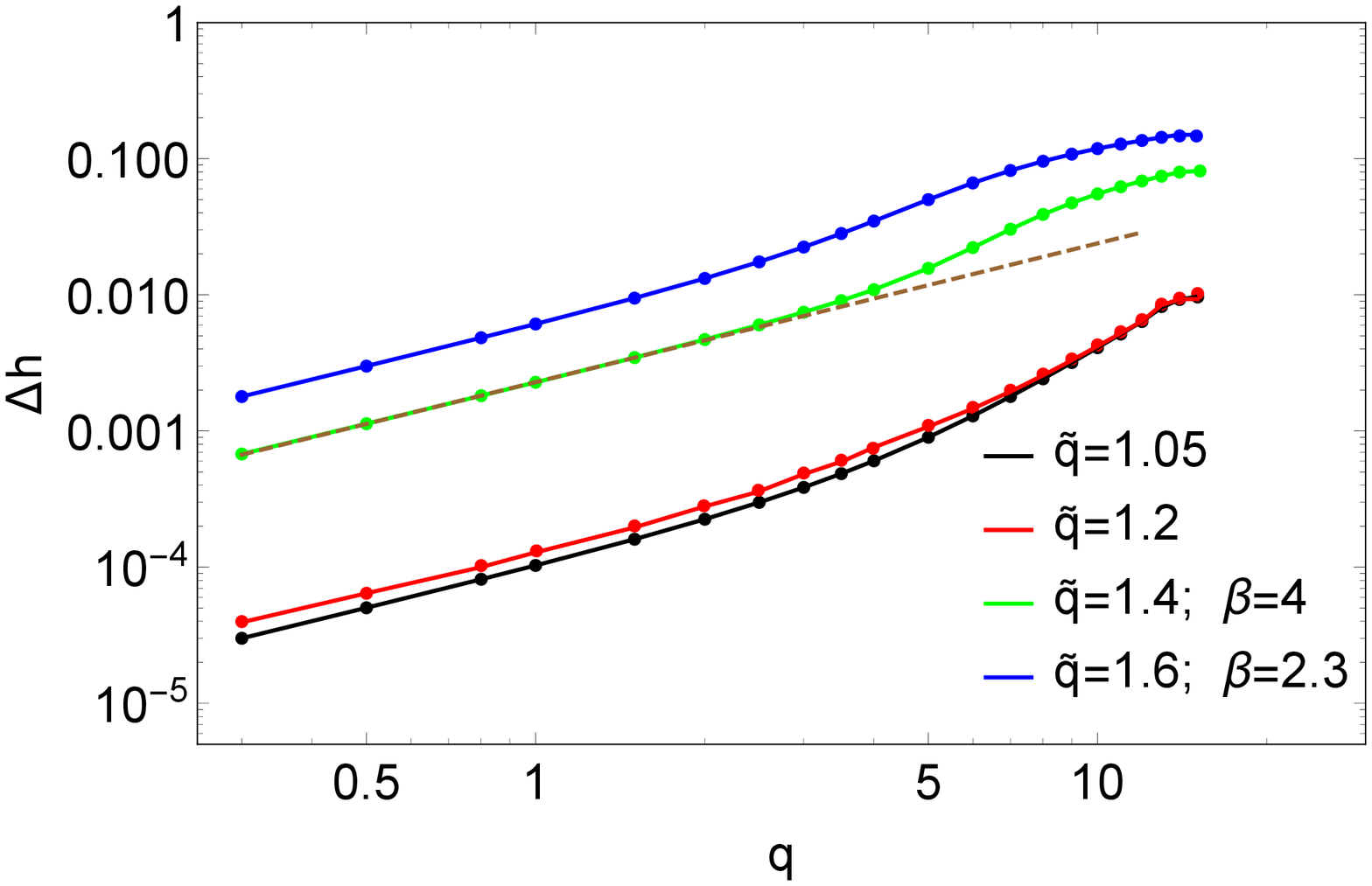}\includegraphics[width=0.4\textwidth, height=0.27 \textwidth]{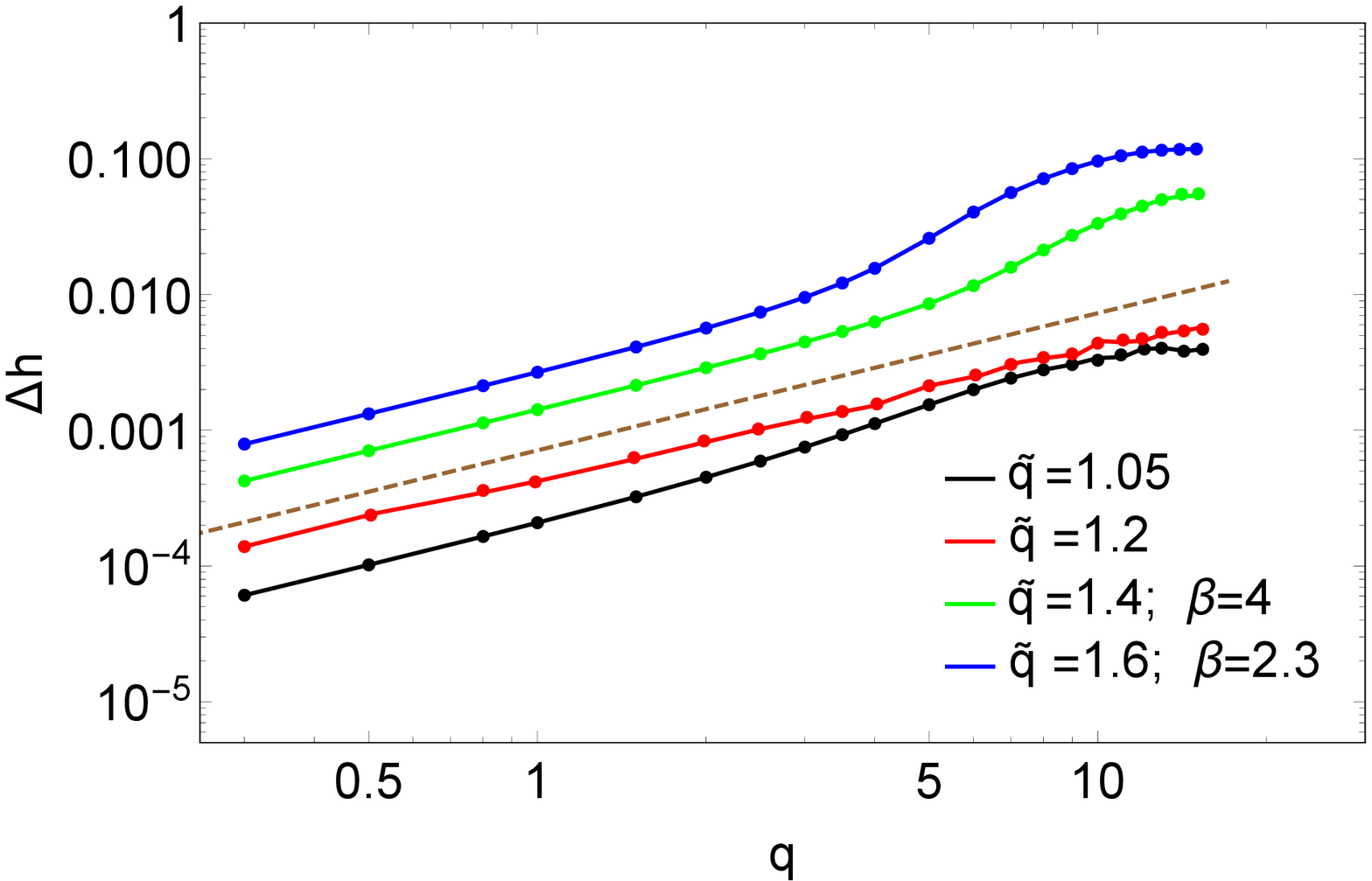}\\
\begin{footnotesize}Levy regime (Symmetric case)~~~~~~~~~~~~~~~~~~~~~~~~~~~~~~~~~Levy regime (Asymmetric case)\end{footnotesize}\\
\includegraphics[width=0.4\textwidth, height=0.27 \textwidth]{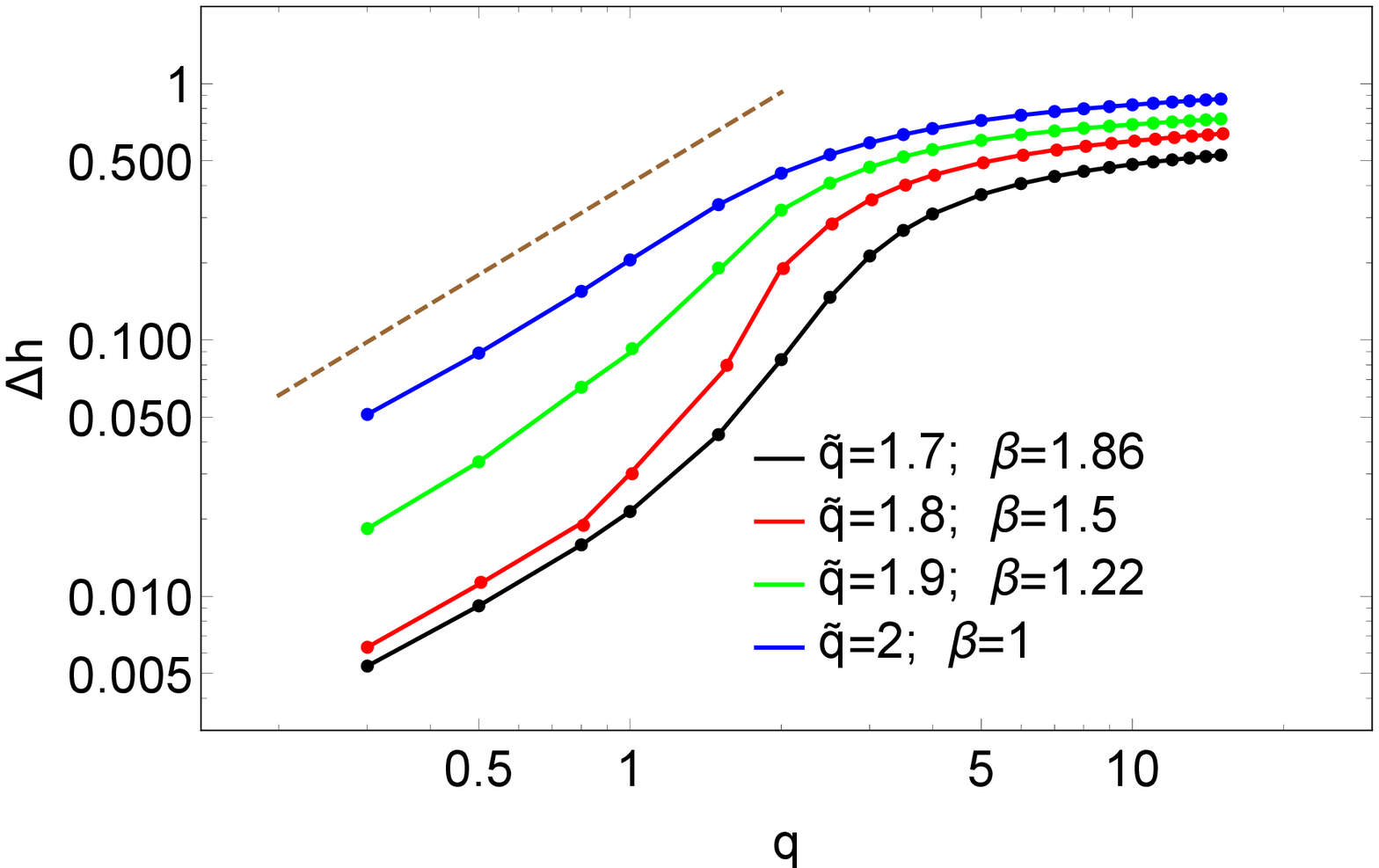}\includegraphics[width=0.4\textwidth, height=0.27 \textwidth]{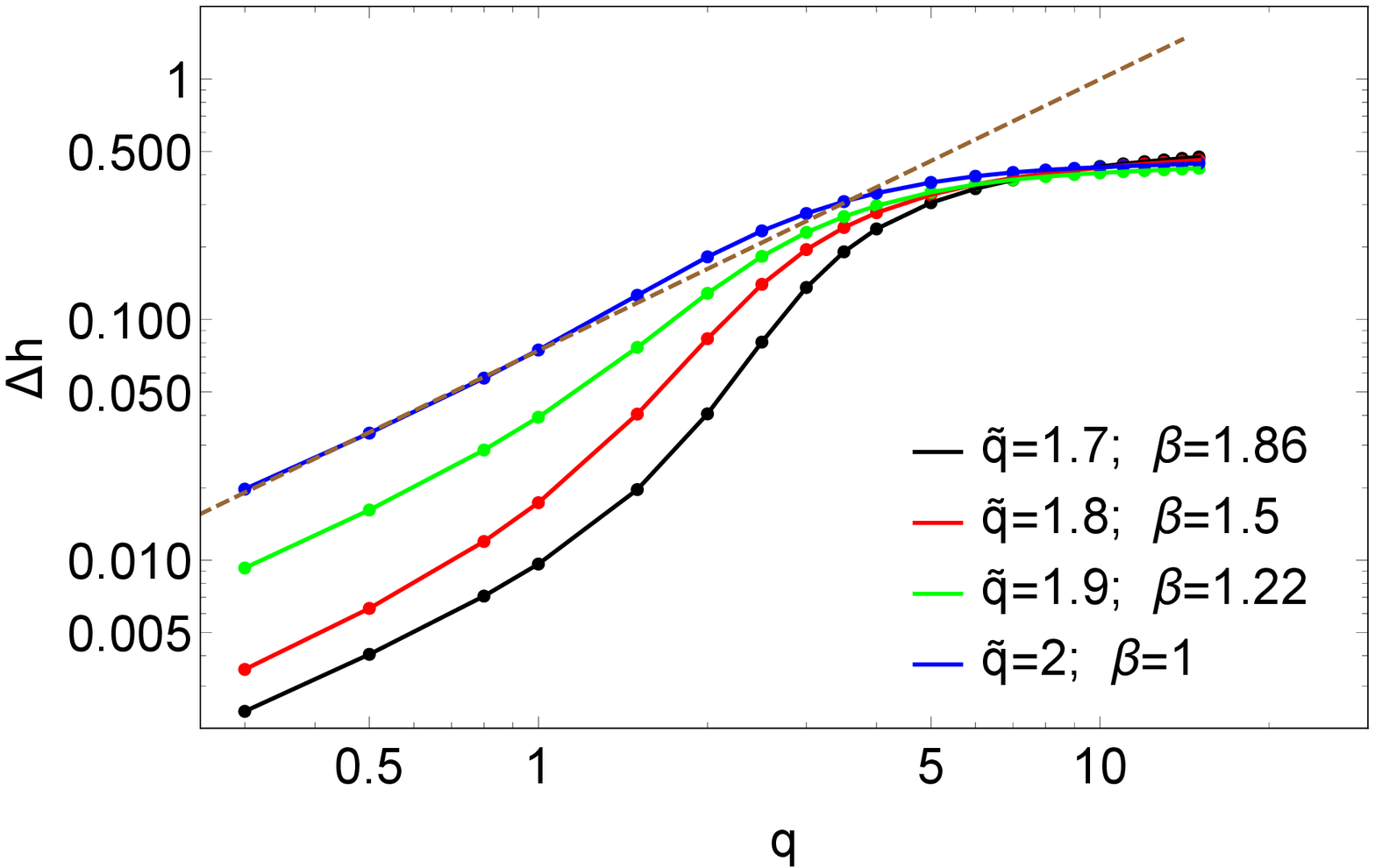}
\end{center}
\begin{footnotesize}
\caption{Spread $\Delta h$ of generalized Hurst exponent calculated for time series of uncorrelated $M=10^{10}$ data in various broad probability distributions originated from Tsallis PDF. Examples of symmetric and asymmetric $\tilde{q}$Gaussians are shown in logarithmic scale. Both regimes, i.e., the Gaussian ($\tilde{q}=1.05, 1.2, 1.4, 1.6$) attractor regime and the Levy attractor regime ($\tilde{q}=1.7, 1.8, 1.9, 2$) are penetrated as labeled on individual panels. The dotted line represent slope corresponding to $\mu=1$ for Gaussian regime and $\mu=1.15$ for Levy regime.}
\label{logdeltah}
\end{footnotesize}
\end{figure}

Fig.~\ref{fig55} shows the main results obtained for variety of fat tailed PDF. From the perspective of many $q$, it can also be seen that the Hurst exponent profile $h(q)$ strictly depends on $\tilde{q}$ (or $\beta$) and the (a)symmetry of distributions. The stronger effect is obviously seen in the Levy attractor regime. The values of $h(q)$ grow much faster for symmetric distributions, especially for small fluctuations of time series, i.e. for $q <0$. For example, if $\beta \approx 1$ $h (-10)=0.68$ for the asymmetric case and  $h(-10)=1$ for the symmetric case respectively.
To make this analysis more exhaustive from quantitative point of view, we have shown the spread $\Delta h$ vs the moment $q$ in logarithmic scale in Fig.~\ref{logdeltah}. This scale more clearly distinguishes several ranges of $q$ parameter for which the apparent multifractal $q$-dependent spread $\Delta h_{FT}$ induced by the presence of fat tails has the particular quantitative form.

Looking first at distributions from Gaussian attractor (top panels of Fig.~\ref{logdeltah}) we see that for small range of deformation parameter $q\leq 3$ dependence $\Delta h_{FT}$ on the maximal moment $q$ used to calculate such spurious multifractal spread may be well approximated by a power law
\begin{equation}
\Delta h_{FT}= C(\beta)q^{\mu}
\label{eq9}
\end{equation}
where $\mu\approx 1$ for all distributions in this attractor.

\begin{table}[]
\begin{footnotesize}
\centering
\label{table1}
\begin{tabular}{|l|l|l|l|l|}
\hline
\multicolumn{5}{|c|}{\textbf{Gaussian regime - symmetric PDF}}                           \\ \hline
$\tilde{q}$                 & 1.05   & 1.2     & 1.4    & 1.6   \\ \hline
$\beta$                     & 39     & 9       & 4      & 2.3   \\ \hline
$C(\beta)(\times 10^{-3})$        & $0.132 $ & $0.154$ & $2.35$ & $6.42$ \\ \hline
$\mu$                       & $1.06\pm0.02$   & $1.06\pm0.02$    & $1.02\pm0.01$   & $1.05\pm0.02$  \\ \hline
$\Delta_{q=15}h_{FT}$ & $9\cdot10^{-3}$  & $9\cdot10^{-3}$   & $8\cdot10^{-2}$   & $1.5\cdot10^{-1}$  \\ \hline
\multicolumn{5}{|c|}{\textbf{Gaussian regime - asymmetric PDF}}                          \\ \hline
$\tilde{q}$                 & 1.05   & 1.2     & 1.4    & 1.6   \\ \hline
$\beta$                     & 39     & 9       & 4      & 2.3   \\ \hline
$C(\beta)(\times 10^{-3})$     & $0.251$ & $0.422$  & $1.46$ & $2.76$   \\ \hline
$\mu$                       & $1.05\pm0.02$   & $0.95\pm0.03$    & $1.01\pm0.02$   & $1.03\pm0.01$  \\ \hline
$\Delta_{q=15}h_{FT}$ & $4\cdot10^{-3}$  & $6\cdot10^{-3}$   & $5\cdot10^{-2}$   & $1.1\cdot10^{-1}$  \\ \hline
\end{tabular}
\caption{Saturated values of spurious multifractal spread $\Delta_{q=15} h_{FT}$ and parameters of  power law fit to $\Delta h_{FT}(q)$ dependence obtained for data in Fig.~\ref{logdeltah} from  various broad probability distributions originated from Tsallis distributions in Gaussian attractor regime for $q\leq 5$. The corresponding $\tilde{q}$ index and $\beta$ decay exponent of PDF tails is shown for completeness. For notation see the main text.}
\label{table1}
\end{footnotesize}
\end{table}

\begin{table}[]
\begin{footnotesize}
\centering
\label{table2}
\begin{tabular}{|l|l|l|l|l|}
\hline
\multicolumn{5}{|c|}{\textbf{Levy regime - symmetric PDF}}                          \\ \hline
$\tilde{q}$              & 1.7   & 1.8     & 1.9    & 2   \\ \hline
$\beta$                     & 1.86     & 1.5       & 1.22      & 1   \\ \hline
$C(\beta)(\times 10^{-2})$     & $2.33$ & $3.26$ & $9.12$ & $19.3$ \\ \hline
$\mu$                       & $1.33\pm0.05$   & $1.33\pm0.05$    & $1.29\pm0.04$   & $1.17\pm0.03$  \\ \hline
$\Delta_{q=15}h_{FT}$ & 0.52  & 0.62   & 0.73   & 0.85  \\ \hline
\multicolumn{5}{|c|}{\textbf{Levy regime - asymmetric PDF}}                           \\ \hline
$\tilde{q}$              & 1.7   & 1.8     & 1.9    & 2   \\ \hline
$\beta$                     & 1.86     & 1.5       & 1.22      & 1   \\ \hline
$C(\beta)(\times 10^{-2})$    & $1.48$ & $2.26$  & $4.21$ & $8.18$   \\ \hline
$\mu$                       & $1.21\pm0.06$   & $1.37\pm0.09$    & $1.25\pm0.04$   & $1.16\pm0.02$  \\ \hline
$\Delta_{q=15}h_{FT}$ & 0.47  & 0.46   & 0.44   & 0.42  \\ \hline
\end{tabular}
\caption{Same as in Table 1 but for distributions in Levy attractor regime  with corresponding $\tilde{q}$ index and $\beta$ decay exponent of PDF tails. The scaling range  $q\leq 1$ was used to fit power law parameters. For notation used see the main text.}
\end{footnotesize}
\end{table}
The corresponding values of $C(\beta)$ coefficients and $\mu$ exponents are collected in Table 1. Note that plots for $\tilde{q}=1.05$ and $1.2$ correspond to PDF close to normal distribution so that shown result recreates in fact the spurious multifractality generated by FSE and is not connected with the influence of discussed effect of broad distribution. Thus only green and blue curves ($\tilde{q}=1.4$ and $\tilde{q}=1.6$) practically describe the investigated effect of fat tails. For $3<q<10$ one has the intermediate region while for $q>10$ the saturation of $\Delta h_{FT}(q)$ starts and ends up with terminal values shown for $q=15$ in Table 1.
 It can be also observed that asymmetric case reveals smaller values of spurious spread $\Delta h_{FT}$ than in case of symmetric broad probability distributions.

The case of broad PDF from Levy attractor regime is different (see bottom panels in Fig.~\ref{logdeltah}). The saturation of multifractal spread $\Delta h_{FT}$ occurs at the same level of $\Delta h_{FT}\backsimeq0.45$ independent on the shape of PDF for asymmetric case. Contrary, for the symmetric case the level of saturation depends on $\beta$ exponent and is given quantitatively as function of $\beta$ in Table 2. For very narrow range of $q\lesssim 1$ corresponding to small fluctuations, the power law dependence as in Eq.~(\ref{eq9}) is fulfilled with related parameters shown also in Table 2. The intermediate region of $\Delta h_{FT}(q)$ is much wider here than for Gaussian regime. Note that the scaling range where the power law dependence of Eq.(\ref{eq9}) occurs becomes wider for thicker tails of PDF, i.e., when $\beta$ increases. Summarizing, the power-law formula of Eq.(\ref{eq9}) and the corresponding saturation values of $\Delta h_{FT}$ offer an easy and immediate way to calculate spurious multifractal effects connected with the presence of broad distribution of data. These general results can be applied to some real data now to reveal to what extend nonlinear multifractal effects are really involved in producing multifractal image of data in time series of any kind.

\section{Application for detecting multifractal components of real financial signals}
In this section we will apply the previous general findings to study multifractal ingredients of real empirical financial data.
In the beginning we shall present an application of findings from the previous section to multifractal analysis of price weighted DJIA index for various time-lags. DJIA belongs to world oldest stock indices connected with the most mature American stock market.

We created unweighted index $I(t)$ based on high frequency price data from the companies listed in Dow Jones Industrial Average (DJIA). The time interval between consecutive records for all companies has been chosen as $\delta t= 5\textrm{sec}$. We considered price returns of 30 companies (AA, AIG, AXP, BA, BAC, CAT, CSCO, CVX, DD, DIS, HD, HPQ, IBM, INTC, JNJ, JPM, KFT, KO, MCD, MMM, MRK, MSFT, PFE, PG, T, TRV, UTX, VZ, WMT and XOM)\footnote{notation according to Bloomberg scheme} collected from the period Jan. 01, 2008 -- July 31, 2011\footnote{data obtained from www.tickdata.com web side}.
For each company 902 days trade was provided, i.e., 705,364 data points (782 price data during the day). The commonly accepted definition of log-returns for time series $I(t_{\delta t})$ representing the index value at time $t_{\delta t}$ was used
$
R \equiv R(t_{\delta t},\Delta t) = \ln I(t_{\delta t} + \Delta t) - \ln I(t_{\delta t}).
$
As another standard procedure, we calculated normalized and centered returns
$x \equiv r(t_{\delta t},\Delta t)$ defined as
$
x = {R - \langle R \rangle_T \over v},
$
where
$
v = (\langle R^2 \rangle_T - \langle R \rangle_T^2)^{1/2 }
$
is the standard deviation of returns over the period $T$ and $\langle \dots \rangle_T$ denotes a time average.
In addition, all overnight returns have been removed, because they cover a much longer time interval introducing unwanted false information on trading. The time $t_{\delta t}$ was assumed to change with the step (tick) $\delta t= 5\textrm{sec}$. This way, one obtains four time series of similar length (approximately 700,000 data points) for arbitrary time-lag $\Delta t$. We used respectively $\Delta t=\{30\textrm{sec},60\textrm{sec},5\textrm{min},10\textrm{min}\}$.

For comparison we also present a systematic study of such characteristics for the Polish stock market index WIG20 (Warszawski Index
Gieldowy - Warsaw Stock Market Index) over the period Nov. 17, 2001 -- Feb. 13, 2018 for the time lags $\Delta t=\{30\textrm{sec},60\textrm{sec},5\textrm{min},10\textrm{min}\}$. This market is commonly classified as still emerging but at least with no doubts much less developed.

\begin{figure}[ht!]
\begin{center}
\includegraphics[width=0.85\textwidth, height=0.5  \textwidth]{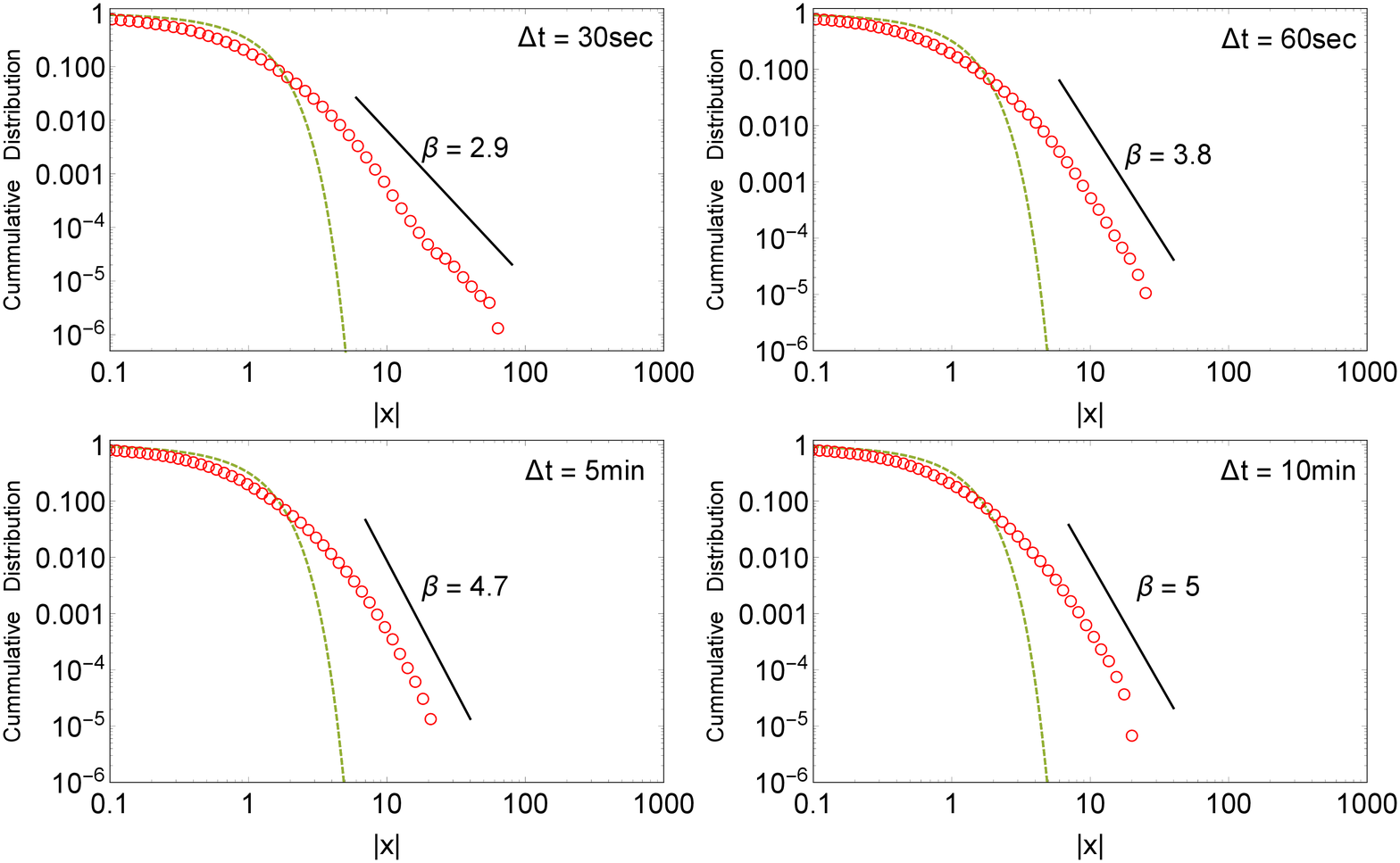}
\end{center}
\caption{Cumulative distributions of moduli of the DJIA normalized logarithmic returns from the period January 1, 2008 -- July 31, 2011 for several time-lags $\Delta t=\{30\textrm{sec},60\textrm{sec},5\textrm{min},10\textrm{min}\}$. The dashed line corresponding to cumulative Gaussian distribution reveals the heavy tailed character of PDF for empirical returns.}
\label{fig7}
\end{figure}

\begin{figure}[ht!]
\begin{center}
\includegraphics[width=0.85\textwidth, height=0.5  \textwidth]{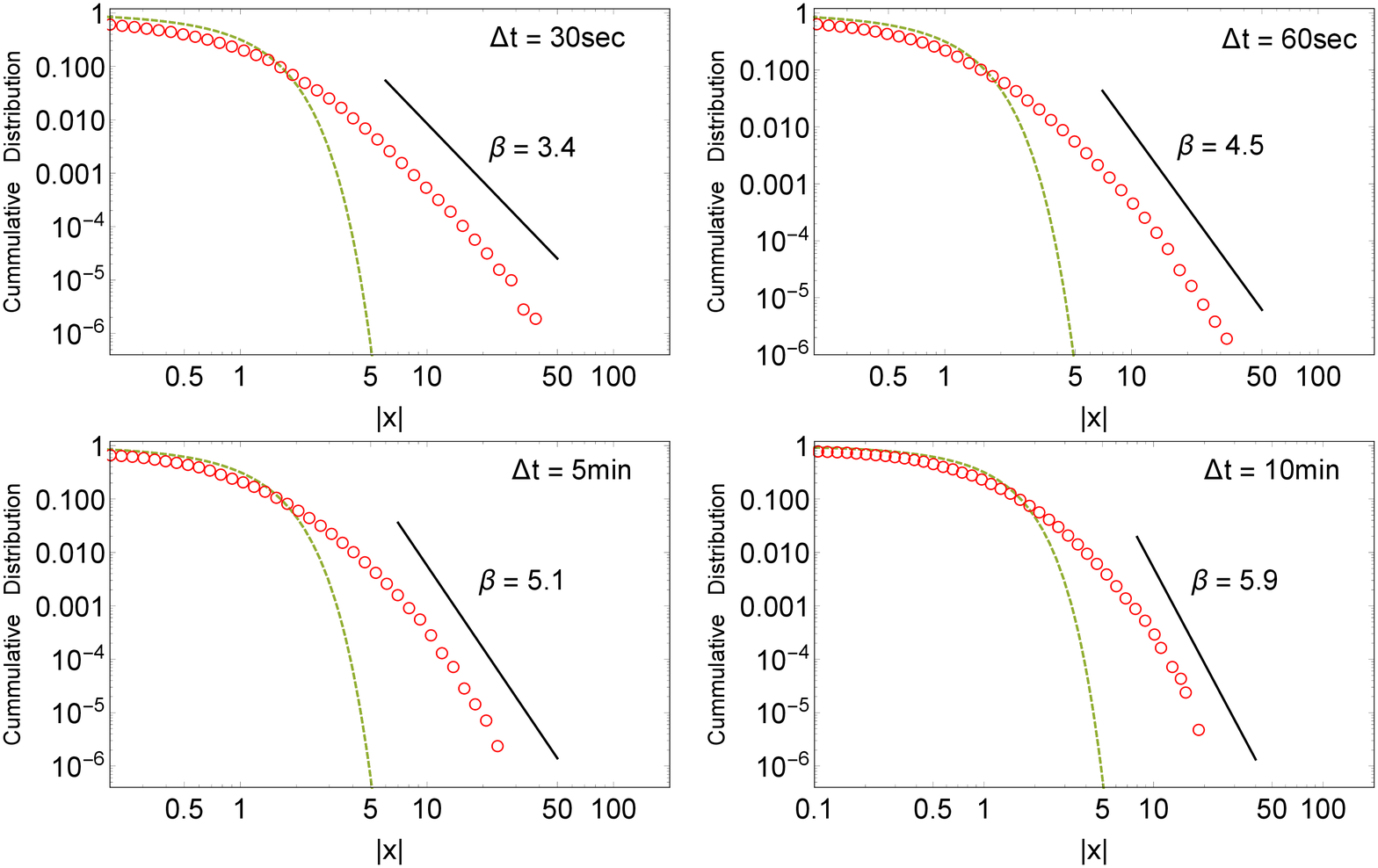}
\end{center}
\caption{The same as in Fig.~\ref{fig7} but for WIG20.}
\label{fig15}
\end{figure}

The cumulative distribution function (CDF) of $\Delta t=\{30\textrm{sec},60\textrm{sec},5\textrm{min},10\textrm{min}\}$ of moduli of DJIA and WIG20 price returns collected from the whole  period specified above is shown in Fig.\ref{fig7} and Fig.\ref{fig15}. We present the distributions of moduli of the returns because the distributions of positive and negative fluctuations turned out to be almost symmetrical.

It can be seen that the tails of all distributions are relatively thick and  vanish according to the power law  $P(r\geq |x|)\sim |x|^{-\beta}$. The fat tails of distributions clearly indicate that the nature of the moduli of logarithmic price returns importantly differs from the Gaussian one. Moreover, from the $\tilde{q}$Gaussians point of view, the obtained scaling exponents of tails (calculated from available data as $\beta=\{2.9,3.8,4.7,5\}$ (for DJIA) and $\beta=\{3.4,4.5,5.1,5.9\}$ (for WIG20) respectively for different time-lags $\Delta t=\{30\textrm{sec},60\textrm{sec},5\textrm{min},10\textrm{min}\}$) indicate that these empirical distributions are unstable and should be located (according to CLT) in the Gaussian attractor regime.
\begin{footnotesize}
\begin{figure}[ht!]
\begin{center}
\includegraphics[width=0.7\textwidth, height=0.45  \textwidth]{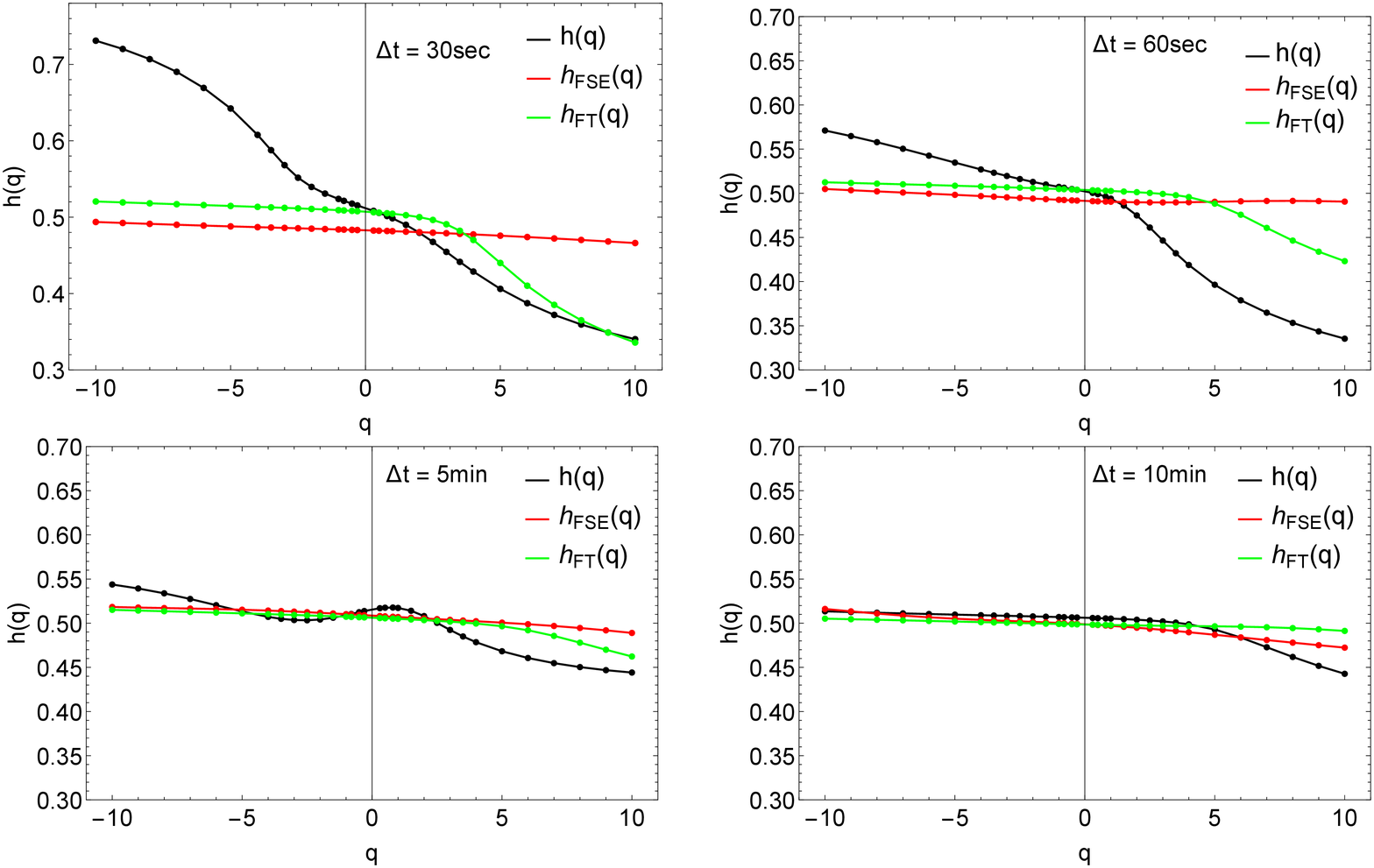}
\includegraphics[width=0.7\textwidth, height=0.45  \textwidth]{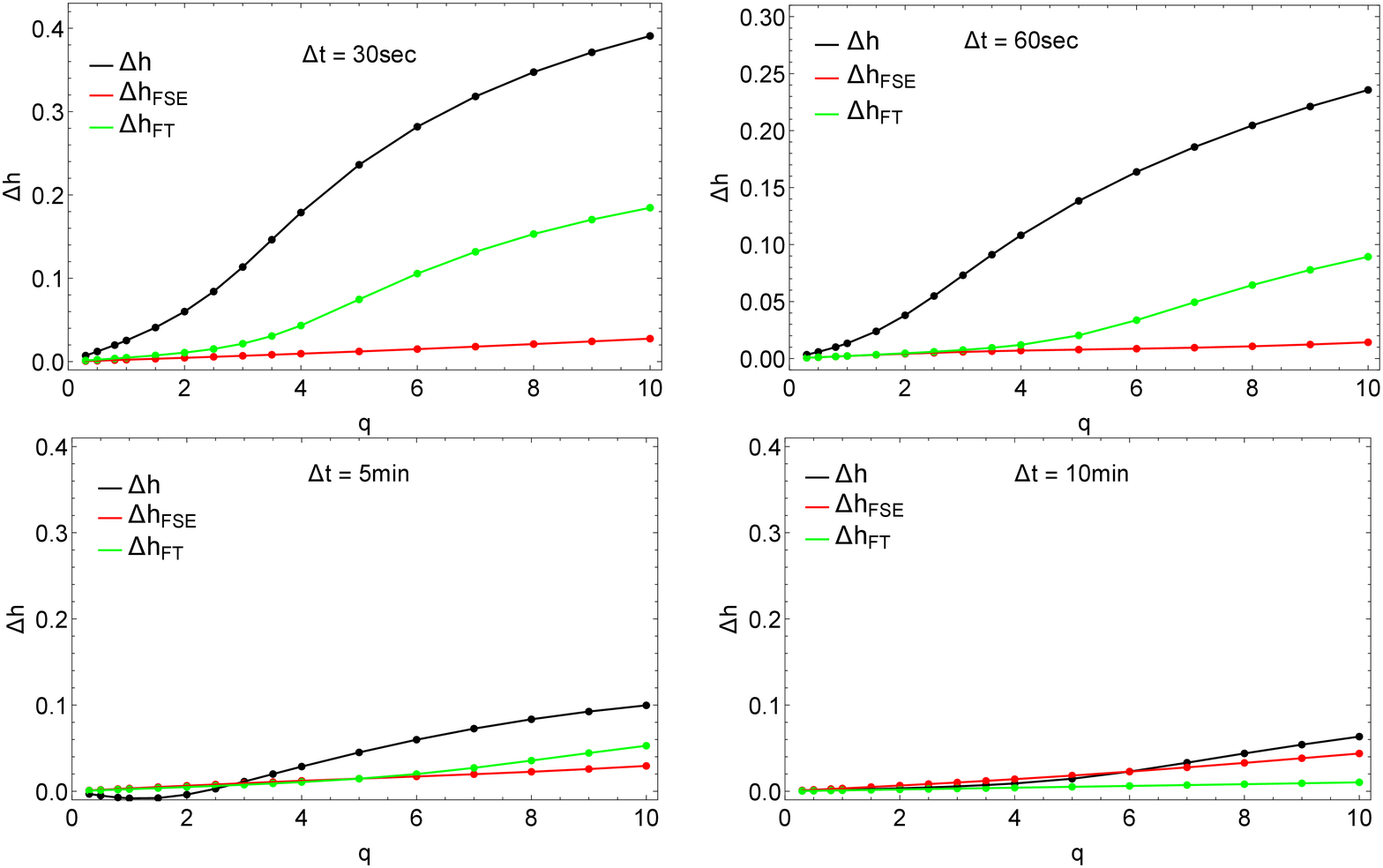}
\end{center}
\caption{The generalized Hurst exponent $h(q)$ (first and second column) and its spread $\Delta h = h(-q)-h(q)$ (third and fourth column) calculated for DJIA logarithmic returns during the period January 1, 2008 -- July 31, 2011 for the time lags $\Delta t=\{30\textrm{sec},60\textrm{sec},5\textrm{min},10\textrm{min}\}$.}
\label{fig8}
\end{figure}
\end{footnotesize}

\begin{footnotesize}
\begin{figure}[ht!]
\begin{center}
\includegraphics[width=0.7\textwidth, height=0.45  \textwidth]{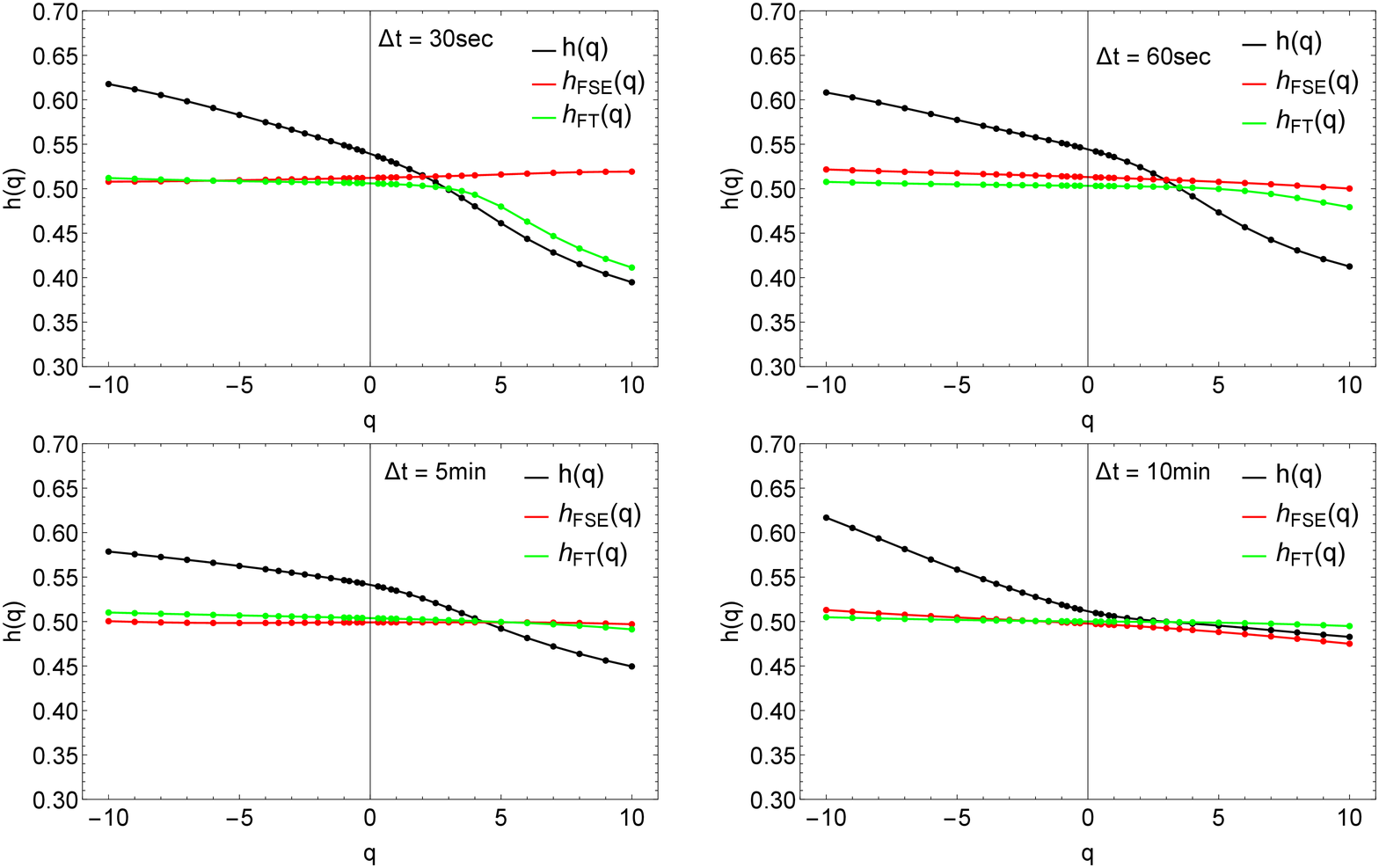}
\includegraphics[width=0.7\textwidth, height=0.45  \textwidth]{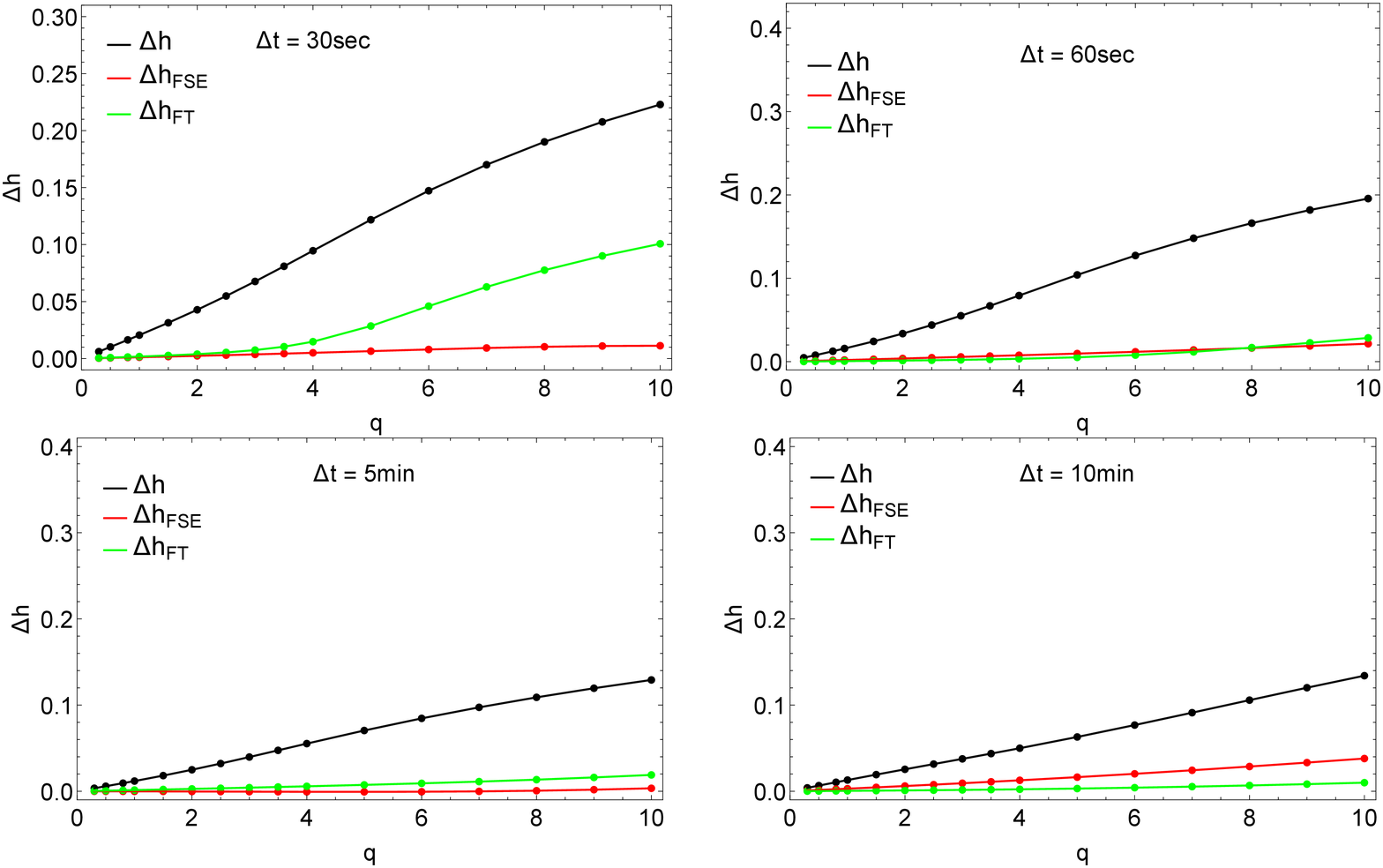}
\end{center}
\caption{The same as in Fig.~\ref{fig8} but for WIG20.}
\label{fig16}
\end{figure}
\end{footnotesize}

In order to check the possible impact of phenomena like: linear and non-linear correlations, FSE and the effect of broad probability distributions on multifractal character of the considered real data for different time-lags, we first calculated the generalized Hurst exponent and its spread $\Delta h$ in a range of moments $-10\leq q \leq10$ for empirical data (time series denoted further on as $s_1$ series). This is indicated in Fig.~\ref{fig8} and Fig.~\ref{fig16} by the black curve.
In the next step, we calculated according to ref.~\cite{63,64} the profile $h(q)_{FSE}$ and the spread $\Delta h_{FSE}$ for the synthetic series of the same length (denoted further on as $s_2$) but drawn from Gaussian distribution and with the same level of linear autocorrelations as empirical series $s_1$. This can be done with the help of Fourier filtering method (see, ref. \cite{ffm} for details). The series $s_2$ contributes to multifractal spectrum only with spurious multifractality related to short length of data (FSE) and to the involved linear autocorrelations.
Its spread $\Delta h_{FSE}$ is given as
\begin{equation}
\Delta h(\xi,M,q)_{FSE}=C_1 M^{-\eta_1}\xi + C_0 M^{-\eta_0} (1-\xi) -  CM^{-\nu}(Q-q)
\end{equation}
where $\xi=2-2H$ and $H=h(q=2)$ is the main Hurst exponent. The values of all parameters in calculations are taken from ref.\cite{63,64}.

Finally, we calculated the spread $\Delta h(q)_{FT}$ of spurious multifractality related to effects of broad data distribution only. This was based on synthetic series (labeled as $s_3$) generated in section 2 in accordance with the $\tilde{q}$Gaussian distribution for the corresponding $\beta$ values of the original empirical distribution.
\begin{figure}[ht!]
\begin{footnotesize}
\begin{center}
\includegraphics[width=0.75\textwidth, height=0.45  \textwidth]{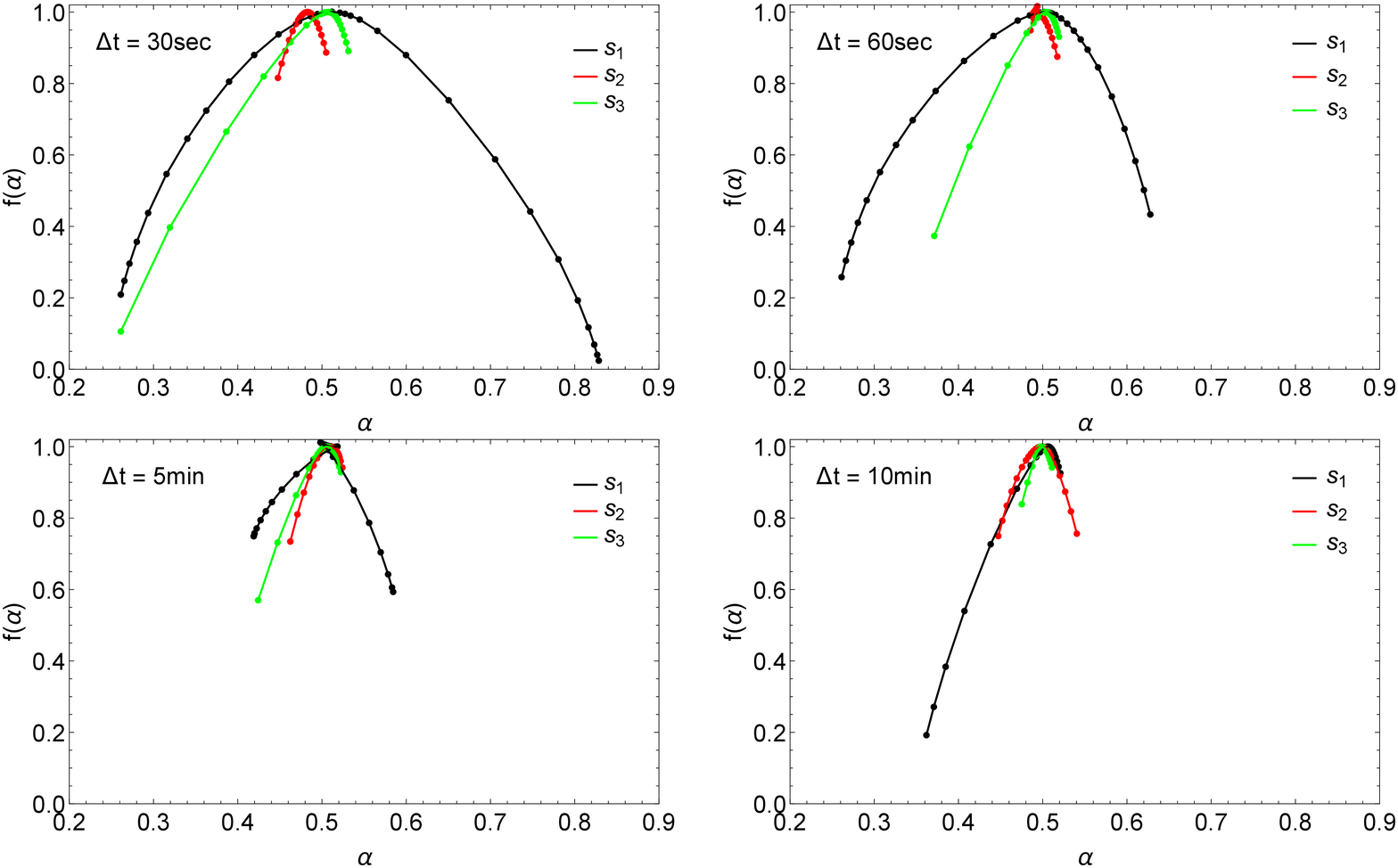}
\end{center}
\caption{Singularity spectra $f(\alpha)$ for DJIA logarithmic returns during the period January 1, 2008 -- July 31, 2011 for the time lags $\Delta t=\{30\textrm{sec},60\textrm{sec},5\textrm{min},10\textrm{min}\}$. The spectra of various spurious multifractal effects contributing to the final picture of observed multifractality are marked with different colors. See the main text for detailed explanations.}
\label{fig9}
\end{footnotesize}
\end{figure}

\begin{figure}[ht!]
\begin{footnotesize}
\begin{center}
\includegraphics[width=0.75\textwidth, height=0.45  \textwidth]{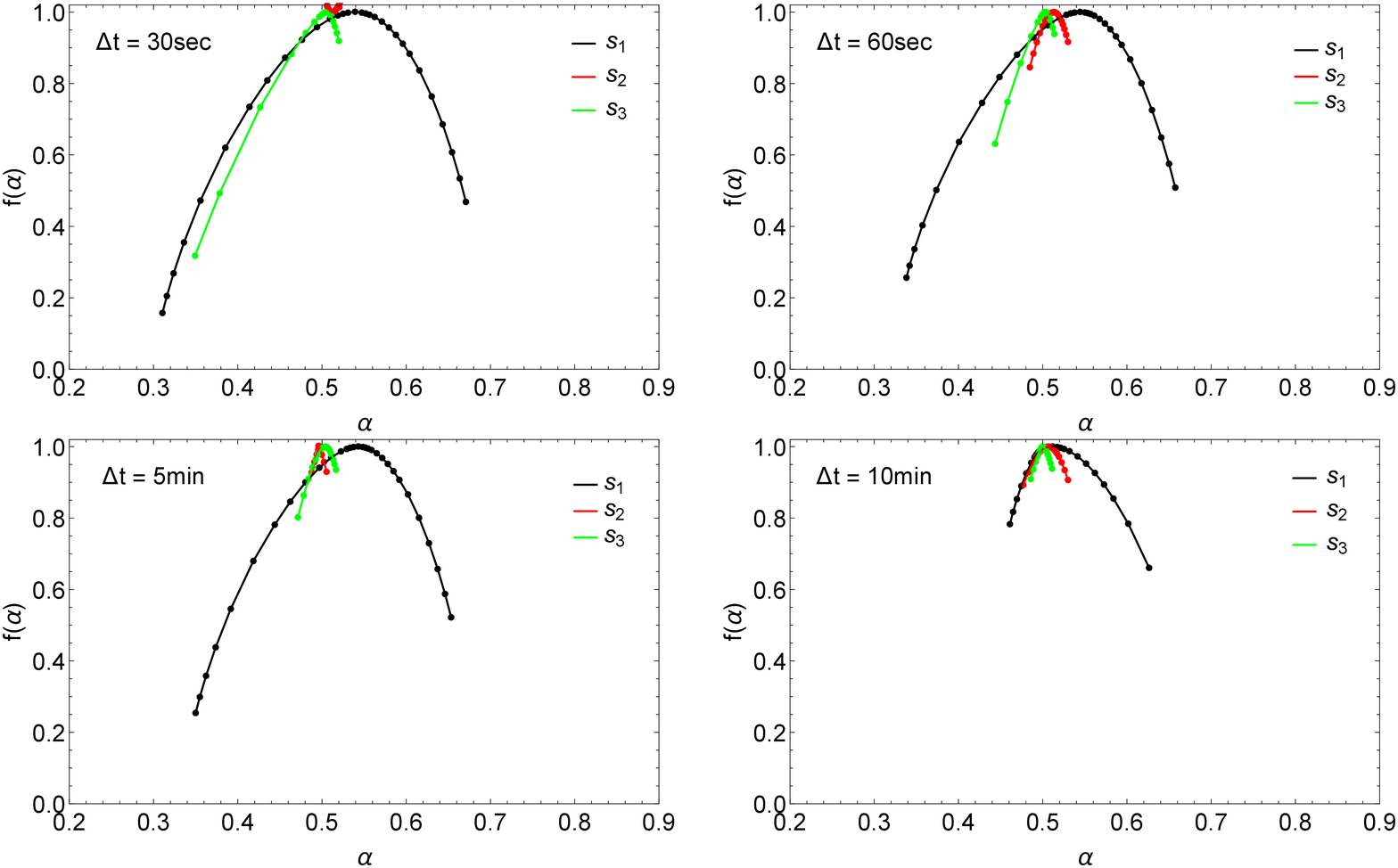}
\end{center}
\caption{The same as in Fig.~\ref{fig9} but for WIG20.}
\label{fig17}
\end{footnotesize}
\end{figure}

The latter series has a probability distribution consistent with real data but neither linear nor non-linear autocorrelations are present in it. We also assumed that the series $s_3$ does not involve the spurious multifractality caused by  FSE because it is much longer ($\sim 10^{10}$ data points) than the real data length and, as already stated, the FSE effects are negligible then ($\Delta h<10^{-3}$).
For all these profiles the multifractal spreads - correspondingly: $\Delta h$, $\Delta h_{FSE}$ and $\Delta h_{FT}$ are shown in last four panels of Fig.~\ref{fig8} and Fig.~\ref{fig16}. Their numerical values have been collected also in Table 3 for various time-lags of DJIA and WIG20 returns. For completeness, the multifractal spectrum $f(\alpha)$ for these cases is also provided in separate Fig.~\ref{fig9} (for DJIA) and Fig.~\ref{fig17} (for WIG20).
\begin{table}[ht!]
\begin{footnotesize}
\centering
\label{table3}
\begin{tabular}{|l|l|l|l|l|}
\hline
           & $\Delta h (\times 10^{-1})$ & $\Delta h_{FSE} (\times 10^{-1})$ & $\Delta h_{FT} (\times 10^{-1})$ & $\Delta h_{NL} (\times 10^{-1})$ \\ \hline
DJIA 30sec & $3.9$  & $0.275$        & $1.84 $       & $1.78 $       \\ \hline
DJIA 60sec & $2.35 $ & $0.142$        & $0.893$       & $1.3$        \\ \hline
DJIA 5min  & $0.99$  & $0.29$         & $0.52$        & $0.26$        \\ \hline
DJIA 10min & $0.63$  & $0.43$         & $0.10$          & $0.09$          \\ \hline
$--------$   & $--------$  & $--------$        & $--------$       & $--------$       \\ \hline
WIG20 30sec & $2.22$  & $0.11$       & $1$       & $1.1$       \\ \hline
WIG20 60sec & $1.95$ & $0.21$        & $0.28$     & $1.46$        \\ \hline
WIG20 5min  & $1.29$  & $0$         & $0.19$      & $1.07$        \\ \hline
WIG20 10min & $1.34$  & $0.38$      & $0.1$    & $0.86$          \\ \hline
$--------$  & $--------$  & $--------$        & $--------$   & $--------$       \\ \hline
EUR 5min   & $2.54$ & $0.27$         & $0.89$        & $1.38$       \\ \hline
GBP 5min   & $3.24$ & $0.32$         & $0.61$        & $2.3$        \\ \hline
RUB 5min   & $4.42$ & $0.27$         & $1.71$       & $2.43$       \\ \hline
\end{tabular}\caption{Summary of multifractal properties of exemplary empirical stock and financial time series. The observable multifractal spread $\Delta h$ is compared quantitatively with spurious constituents $\Delta h_{FSE}$ and $\Delta h_{FT}$. The "true" multifractal content of these series $\Delta h_{NL}$ found only after subtracting its spurious constituents is shown in last column.}
\end{footnotesize}
\end{table}

\begin{figure}[ht!]
\begin{center}
\includegraphics[width=0.75\textwidth, height=0.45  \textwidth]{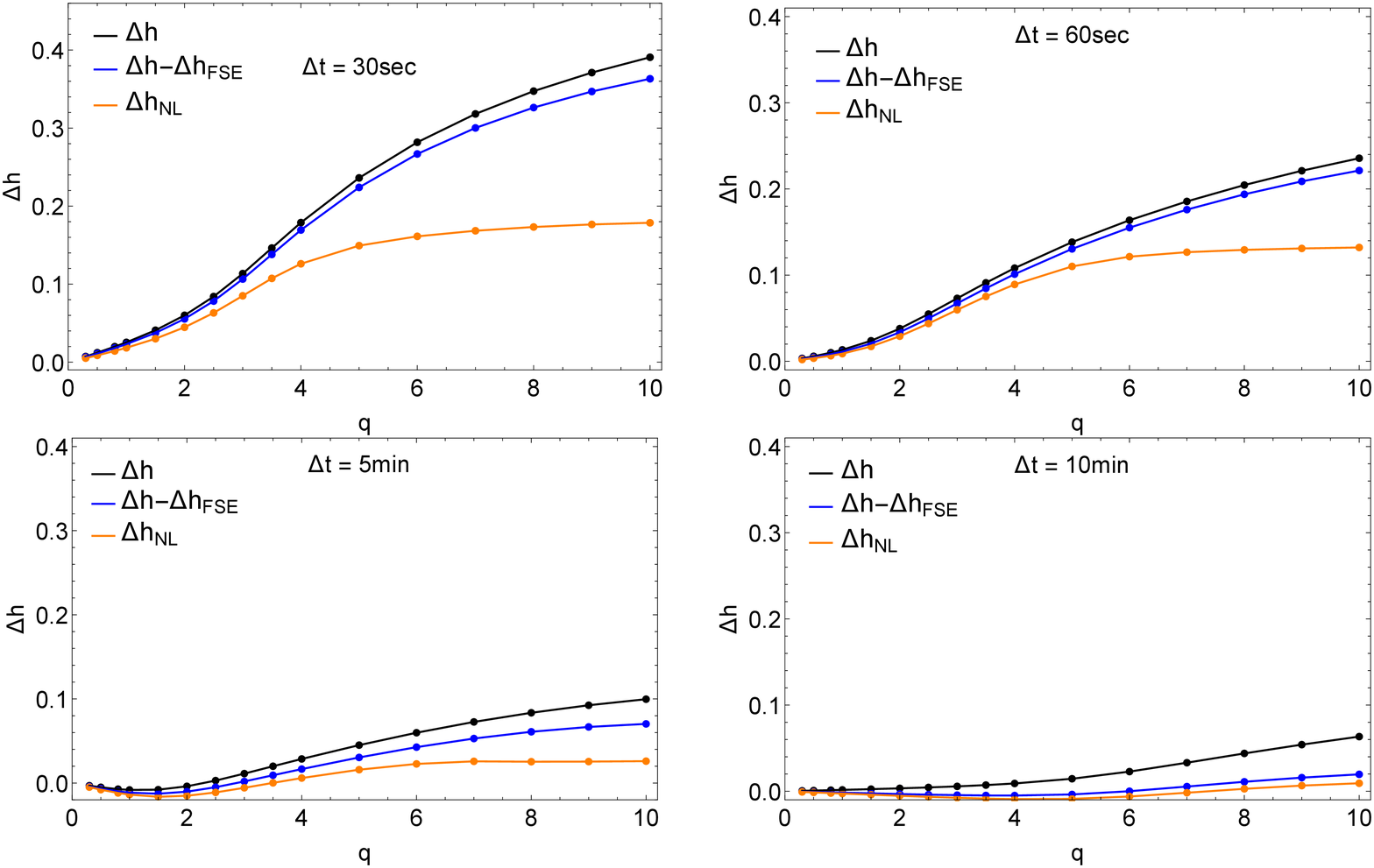}
\end{center}
\caption{The observable multifractal spread $\Delta h$ of DJIA index for various time-lags as a function of deformation parameter $q$. The results are compared quantitatively with its "true" multifractal content $\Delta h_{NL}$ found after subtracting spurious constituents, i.e., $\Delta h_{FSE}$ and $\Delta h_{FT}$. }
\label{fig10}
\end{figure}

\begin{figure}[ht!]
\begin{center}
\includegraphics[width=0.75\textwidth, height=0.45  \textwidth]{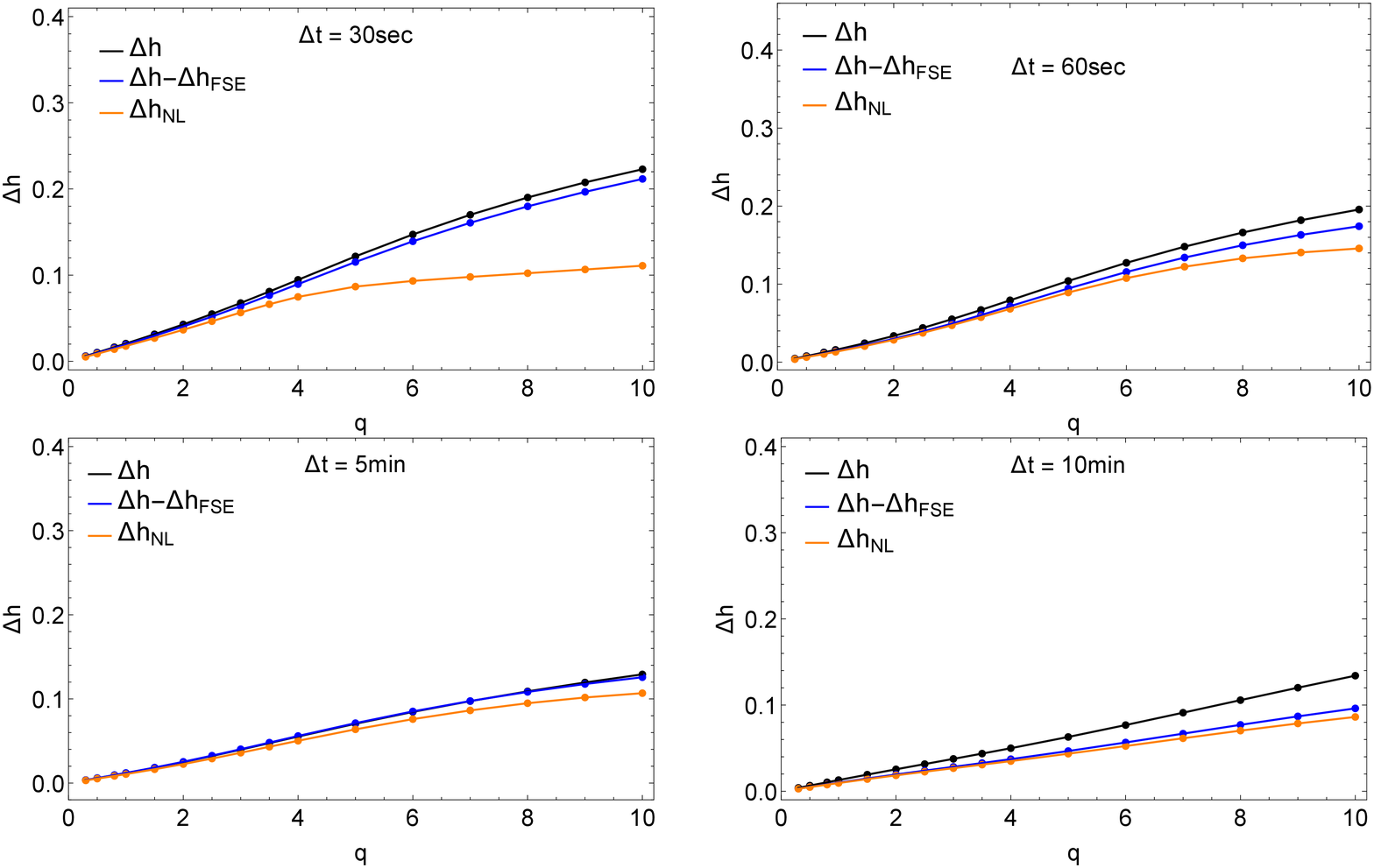}
\end{center}
\caption{The same as in Fig.~\ref{fig10} but for WIG20.}
\label{fig18}
\end{figure}

Having this in mind one can evaluate the upper threshold of multifractal effects associated only with all spurious ingredients and finally the "true" multifractality related only with non-linear correlations which is the most interesting. These results are provided in Fig.\ref{fig10} (for DJIA) and Fig.\ref{fig18} (for WIG20).

Apart from the initial multifractal spread $\Delta h(q)$ made for empirical data within MFDFA, we show in this figure also the "true" multifractal spread for these data $\Delta h_{NL}$. The latter one corresponds to multifractal content of examined series after all spurious effects induced by FSE or by broad distribution of data are subtracted. Thus the $orange~line$ in there, corresponding only to existence of nonlinear correlations in examined series, is obtained as $\Delta h_{NL} = \Delta h(q) - (\Delta h_{FSE}(q) + \Delta h_{FT}(q))$. The multifractal unbiased spread $\Delta h_{NL}$ is shown also and compared with initial value $\Delta h(q)$ and the spurious multifractal effects $\Delta h_{FSE}(q)$ and $\Delta h_{FT}(q)$ separately in Table 3.\\

\begin{figure}[ht!]
\begin{center}
\includegraphics[width=0.42\textwidth, height=0.27  \textwidth]{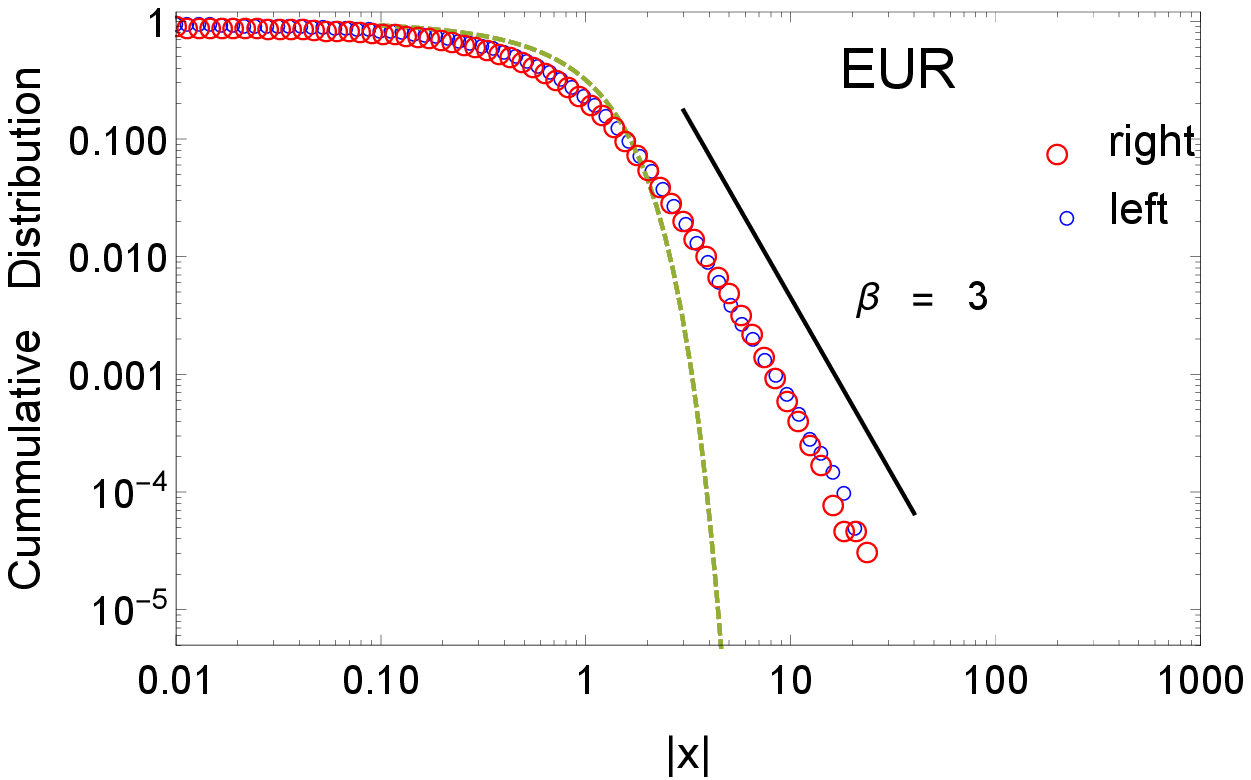}~~~~\includegraphics[width=0.42\textwidth, height=0.27  \textwidth]{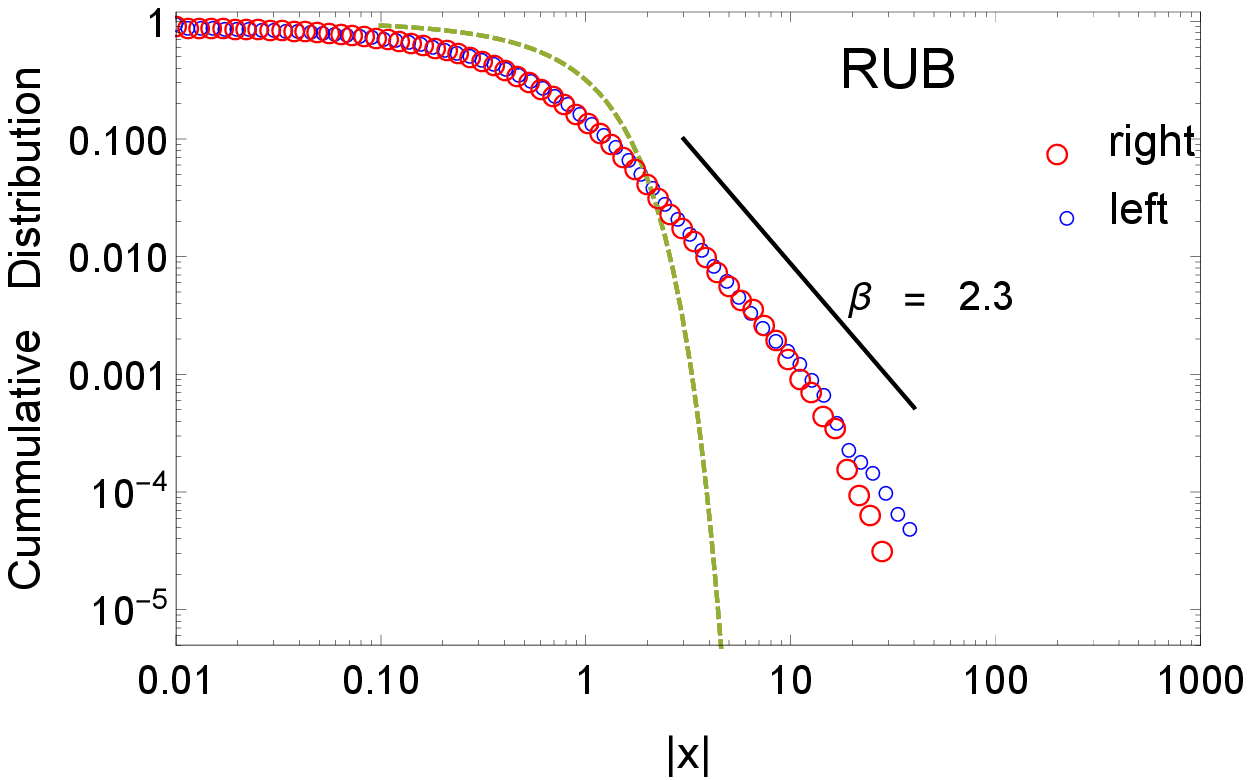}
\includegraphics[width=0.42\textwidth, height=0.27  \textwidth]{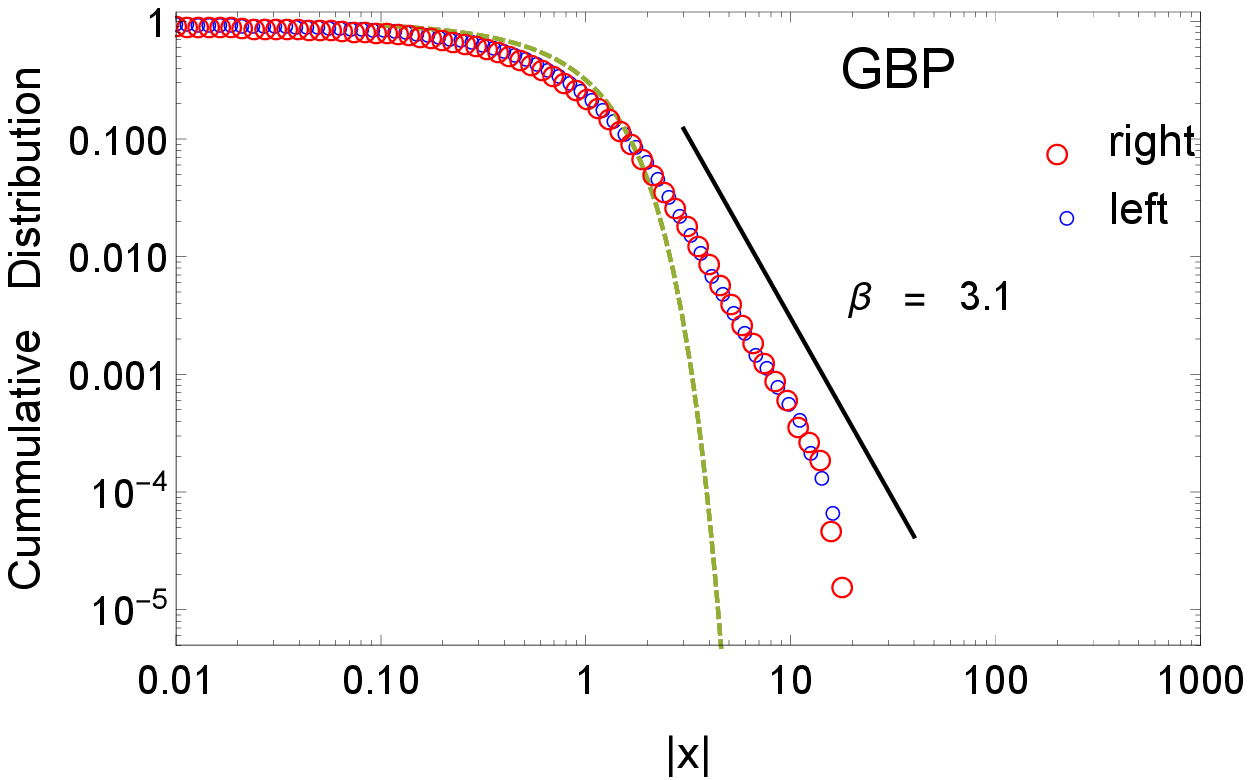}
\end{center}
\caption{Cumulative distributions of moduli normalized returns of EUR/USD, GBP/USDM and RUB/USD from the period January 1, 2014 -- December 31, 2016 for time-lag $\Delta t=5\textrm{min}$. The dashed line corresponds to cumulative Gaussian distribution.}
\label{fig11}
\end{figure}

\begin{footnotesize}
\begin{figure}[h!]
\begin{center}
\includegraphics[width=0.75\textwidth, height=0.45  \textwidth]{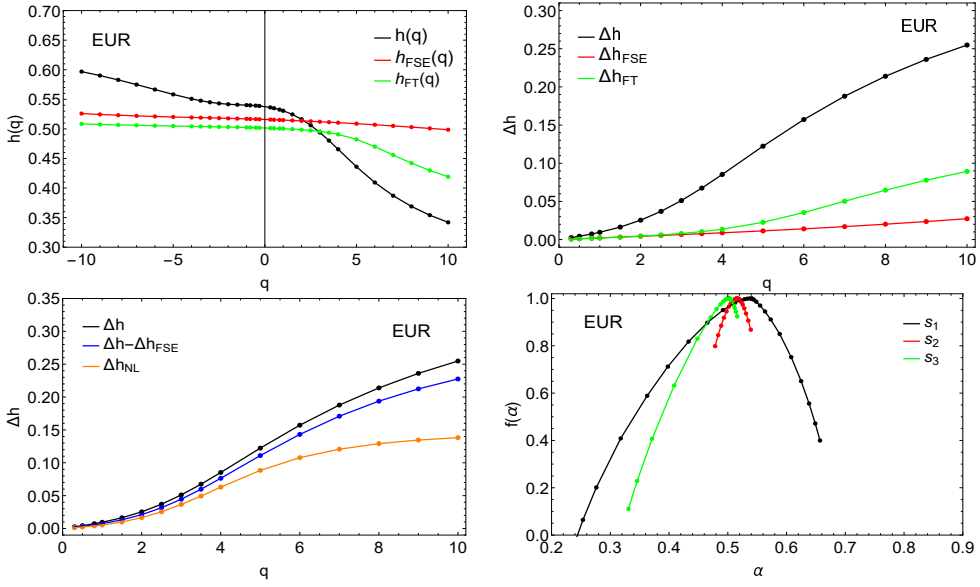}
\end{center}
\caption{The generalized Hurst exponent $h(q)$ (top left), its spread $\Delta h$ (top right) and the "true" multifractal content $\Delta h_{NL}$ found after subtracting spurious constituents, i.e., $\Delta h_{FSE}$ and $\Delta h_{FT}$ (bottom left) calculated for returns fluctuation of EUR/USD over the period January 1, 2014 -- December 31, 2016 for time-lag $\Delta t=5\textrm{min}$. The bottom left panel represents singularity spectra $f(\alpha)$ of various spurious multifractal effects contributing to the final picture of observed multifractality.}
\label{fig12}
\end{figure}
\end{footnotesize}

\begin{figure}[ht!]
\begin{center}
\includegraphics[width=0.75\textwidth, height=0.45  \textwidth]{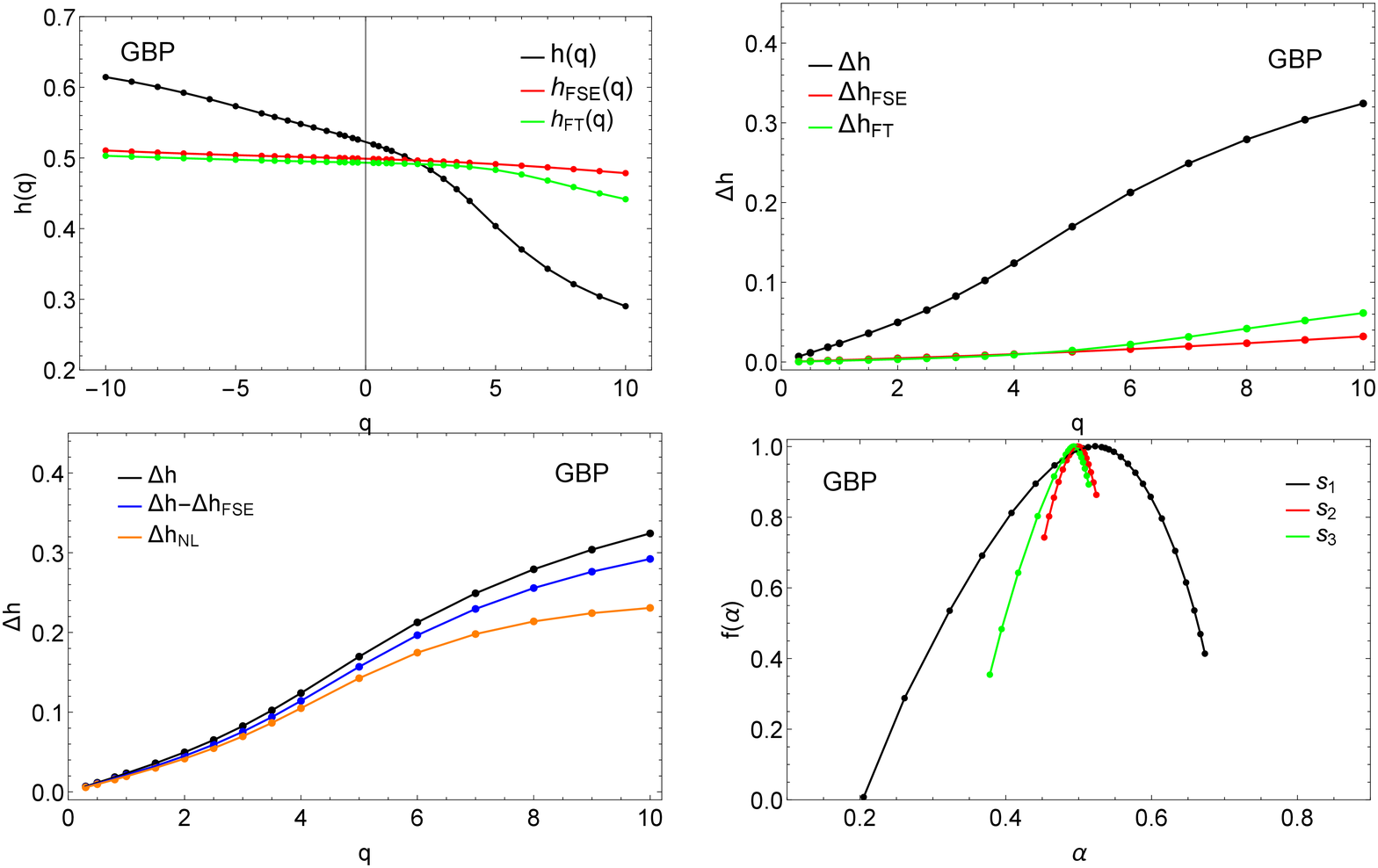}
\end{center}
\caption{The same as in Fig.\ref{fig12} but for GBP/USD.}
\label{fig13}
\end{figure}

It is worth mentioning that the tails of PDF of absolute returns for WIG20 are found thinner and more deformed than in the case of DJIA -- in particular for the minute time scale.
This, in turn, influences $\Delta h_{FT}$ values which are smaller than corresponding ones for DJIA index. Nevertheless the 'true' multifractal content caused by nonlinear correlations $\Delta h_{NL}$ is still more noticeable for WIG20 for the minute time-lags as shown in Table 3.

\begin{figure}[ht!]
\begin{center}
\includegraphics[width=0.75\textwidth, height=0.45  \textwidth]{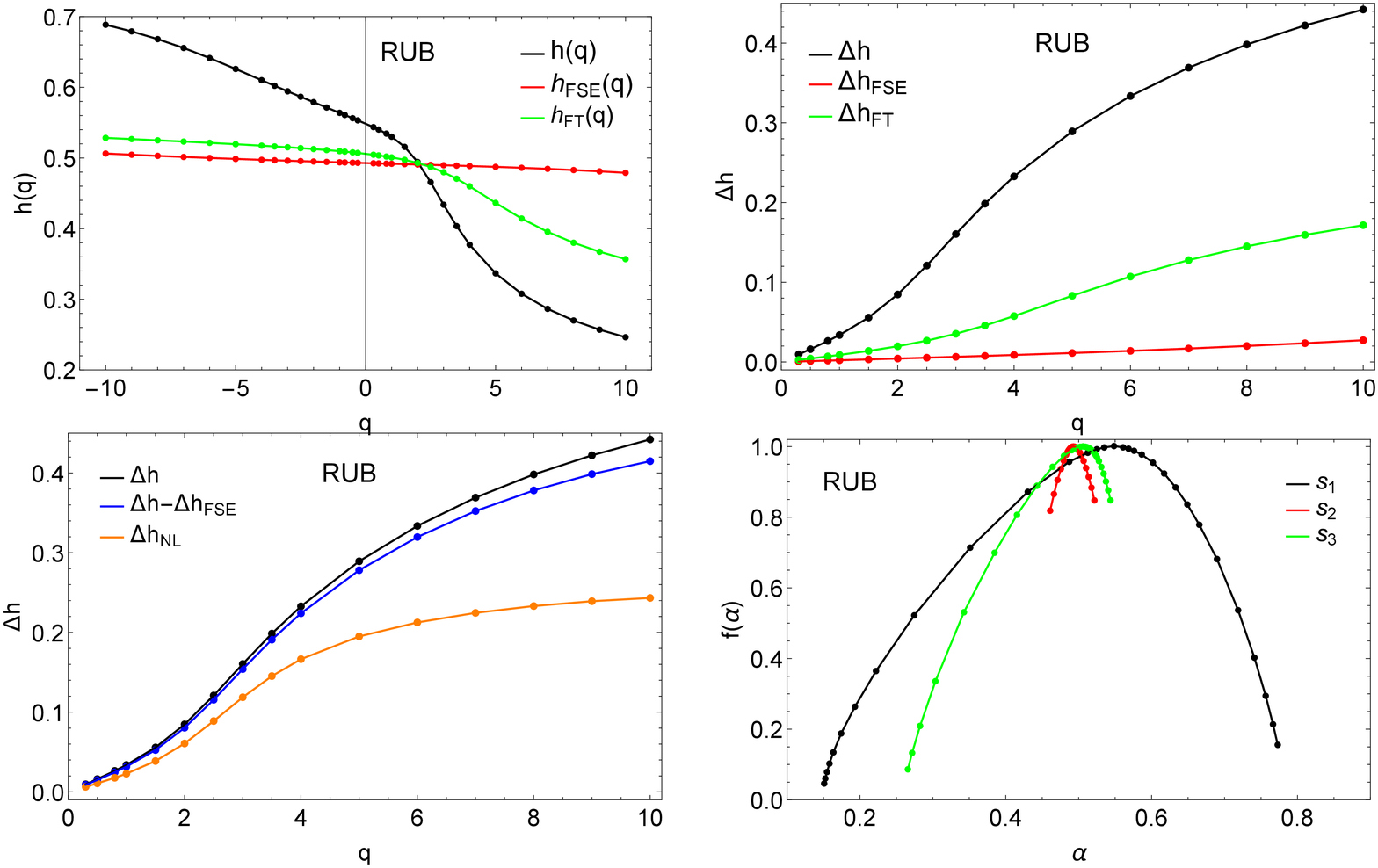}
\end{center}
\caption{The same as in Fig.\ref{fig12} but for RUB/USD.}
\label{fig14}
\end{figure}

%%%%%%%%%%%%%%%%%%%%%%%%%%%%%
As a final example we performed the similar analysis for other kind of data taking the exchange ratios of three currencies from Forex with respect to USD: EUR/USD, GBP/USD and RUB/USD in the period 01.01.2014--31.12.2016 (with the trading hours 8:00 - 22:00) with time-lags of exchange returns $\Delta t= 5$ min ($M=115,736$ data points)\footnote{data downloaded from the Metatrader Platform www.metatrader5.com/}. The results are revealed in similar form as for DJIA and WIG20 index in Figs.~{\ref{fig11}--\ref{fig14}} and provided also in Table 3. The most heavy tail of fluctuation distribution (Fig.\ref{fig11}) is observed for RUB/USD ($\beta=2.3$) presumably related to high level of speculations  involved in trading and manifesting as nonlinear correlations between data. The distributions for the other two exchange rates (EUR/USD and GBP/USD) reveal scaling $P(r\geq |x|)\sim |x|^{-\beta}$ consistent with the known inverse cubic power law ($\beta=3$).

In all considered cases the 'true' multifractal content of data connected with nonlinear properties is seriously reduced with respect to naive description before the spurious effects are subtracted (compare the first and the last column in Table 3). The highest relative influence of spurious effect of heavy tail in PDF on observed multifractal content is observed for RUB/USD exchange rates ($\Delta h_{FT}/\Delta h \sim 40\%$).
$$
$$

\newpage
\newpage
\newpage
$$
$$

\section{Concluding remarks}
The goal of this article was the detailed quantitative analysis of spurious multifractal effects induced by the presence of broad distribution of data in time series. To make this analysis more close to practical application we analyzed PDF with heavy tails generated by a family of $\tilde{q}$Gaussian distribution. They offer to control not only the shape of PDF tails but also the parametric description of the whole distribution - also in its head part. This way, the level of difference with respect to normal or Levy distribution is constantly monitored. Such distributions have their theoretical background in non-extensive statistical physics and are well confirmed to describe many phenomena in majority of complex systems, independently on the particular mechanism of information flow standing behind or the nature of complexity. The $\tilde{q}$Gaussian distribution offers an easy way to pass through all intermediate steps of heavy tailed PDF -- from those with infinite variance and infinite higher statistical moments (lying in so called Levy attractor regime) up to those with finite variance although still with sufficiently heavy tails. The latter case of PDF, according to Central Limit Theorem, drops into Gaussian regime attractor and it seems to be of greatest practical importance particularly in finance because of risk analysis possibility.

This paper has been divided into two major and in some way related parts. The first part concerned the analysis of synthetic data generated from $\tilde{q}$Gaussians distributions. In this part we have explored quantitatively the impact of heavy tailed symmetric and asymmetric probability distributions on multifractality. We assumed that the asymmetric distribution is one for which the right tail (positive fluctuations) has a normal distribution. We observed, among others significantly slower growth (in the Levy regime) of $\alpha_{max}\sim \beta^{-\gamma}$ value, i.e., $\gamma=-0.37$ in the case of asymmetric distributions,  while for symmetric distributions (according to theoretical prediction) $\gamma=1$.  The smaller value of $\gamma$ exponent significantly depletes the multifractal nature of the analyzed data - the decrease of the multifractal spectrum $f(\alpha)$ to the right is much slower than in symmetric case.

Regarding analysis of the whole multifractal spectrum of data from broad distributions we showed that the spurious multifractal effects induced by this kind of PDF can be well described quantitatively by a power law linking the spurious multifractal spread $\Delta h_{FT}=C(\beta)q^{\mu}$ expressed in Hurst language with the multifractal deformation parameter $q$ and the rate of PDF tail decay $\beta$. This power law is well satisfied in the range $-5\leq q\leq 5$ often used by many authors investigating multifractal phenomena in empirical time series. The corresponding $C(\beta)$ and $\mu$ values have been classified by us. We noticed also that the effect of spurious multifractality caused by heavy tails in PDF seems to saturate at the level $q\sim 10\div 15$ due to numerical reasons. These saturated maximal values of the spread $\Delta h_{FT}$ expected to occur in heavy tailed data has also been calculated and presented as a function of $\beta$ exponent shaping PDF tails. These general findings may serve as complementary basis for estimation of all "false" signal of multifractality actually appearing in data of monofractal nature.
Thanks to general considerations conducted in literature so far, enriched with new quantitative findings in this article, one is able to clearly separate three basic ingredients of observable multifractal content into parts generated by: short length of data series and linear correlations being involved ($\Delta h_{FSE}$), effect of heavy tails in PDF ($\Delta h_{FT}$) and nonlinear correlations changing with time scale ($\Delta h_{NL}$). We argued that only the latter one describes the "true" multifractal effect in data of any kind and therefore it should be always clearly separated from other multifractal effects having the spurious character of an false signal in any applicative study.

Going in this direction we finally provided examples from stock market (DJIA and WIG20) and money market (Forex) indicating the real multifractal content of empirical signal in time series against its spurious constituents.
The main conclusion to be drawn from the real data analysis presented in the second part of the paper is that the spurious effect of fat tails of distributions seems to be tremendously much more important than spurious multifractal effects of finite data length or the effect caused by linear correlations involved in series.
Looking at details revealed in Table 3 one may expect that the true multifractal content of empirical data ($\Delta h_{NL}$) might be only around half of the observable expectation level (i.e., $\Delta h$) measured initially within MFDFA. The spurious multifractality is particularly present in returns calculated for short time-lags since in this case the tails of distributions become more thick -- the effect is particularly visible for the currencies considered here (EUR/USD, GBP/USDM, RUB/USD) and for DJIA index for the time scales $\Delta t=30\textrm{sec}$ and $60\textrm{sec}$. Interestingly, the stronger nonlinear effects are visible for WIG20 index especially for longer time scales $\Delta t=5\textrm{min}$ and $10\textrm{min}$. This can be caused by the fact that the Polish stock market is much smaller and younger than the US one and there are significantly fewer transactions on Polish market in a given time unit -- in other words the transaction time seems to run relatively 'slower' comparing with more mature markets.

Therefore, one should be very careful drawing conclusions from the multifractal
analysis and interpretation of the observed multifractal spread in any complex system.
In particular, from a practical point of view, the effects that we quantitatively described can have applications in modeling and forecasting the widely understood stock market data.
$$
$$
{\textbf{Acknowledgement}}

This work was partially supported by the Centre for Innovation and Transfer of Natural Sciences and Engineering Knowledge (University of Rzesz\'ow).


\begin{thebibliography}{999}


\bibitem{mf12} S. Ghashghaie, W. Breymann, J. Peinke, P. Talkner, and Y. Dodge, Nature 381, 767 (1996).
\bibitem{mf13} R. N. Mantegna and H. E. Stanley, Nature 383, 587 (1996).
\bibitem{mf14} B. B. Mandelbrot, Sci. Am. 298, 70 (1999).
\bibitem{kantelhardt-preprint} J.W. Kantelhardt, arXiv: 0804.0747v1 [phys.data-an].
\bibitem{mf1} Z. Eisler, J. Kert{\'e}sz, Physica A 343 (2004) 603.
\bibitem{mf2} H. G. E. Hentschel, I. Procaccia, Physica D 8 (1983) 435.
\bibitem{mf3} T. C Halsey, M.H. Jensen, L. P. Kadanoff, I. Procaccia, B. I. Shraiman, Phys. Rev. A 33 (1983) 1141.



\bibitem{18} M. H. Jensen, L.P. Kadanoff, A. Libchaber, I. Procaccia, J. Stavans, Phys. Rev. Lett. 55 (1985) 2798.
\bibitem{19} J. F. Muzy, E. Bacry, A. Arneodo, Phys. Rev. Lett. 67 (1991) 3515.
\bibitem{20} F. S. Labini, M. Montouri, L. Pietronero, J. Physique IV 8 (1998) Pr-115 Pr-118.
\bibitem{21} K. Ivanova, H.N. Shirer, E. E. Clothiaux, N. Kitova, M. A. Mikhalev, T. P. Ackerman, M. Ausloos, Physica A 308 (2002) 518.
\bibitem{22} Y. Ashkenazy, D. R. Baker, H. Gildor, S. Havlin, Geophys. Res. Lett. 30 (2003) 2146.
\bibitem{22a} J. de Souza, SM Duarte Queiro's, AM Grimm,  Chaos (Woodbury, N.Y.) 23 (2013) 023130.
\bibitem{23} P. C. Ivanov, L. A. N. Amaral, A. L. Goldberger, S. Havlin, M. G. Rosenblum, Z. R. Struzik, H. E. Stanley, Nature 399 (1999) 461.
\bibitem{24} M. Ausloos, Translation effects, Chaos, Solitons Fractals 45 (2012) 1349.
\bibitem{24a} S. Dro\.{z}d\.{z}, P. O\'{s}wi\c{e}cimka, A. Kulig, J. Kwapie\'{n}, K. Bazarnik, I. Grabska-Gradzi\'{n}ska, J. Rybicki, and M. Stanuszek, Information Sciences 331 (2016) 32.
\bibitem{24a1} R. Rak, S. Bwanakare, Acta Phisica Polonica A 129(5) (2016) 922.
\bibitem{24b}  A. Kulig, J. Kwapie\'{n}, T.Stanisz, S. Dro\.{z}d\.{z}, Information Sciences 375 (2017) 98.
\bibitem{25} C. Amitrano, A. Coniglio, P. Meakin, M. Zanetti, Phys. Rev. B 44 (1987) 4974.
\bibitem{26} H.E. Stanley, P. Meakin, Nature 335 (1988) 405.
\bibitem{27} A. Fisher, L. Calvet, B. Mandelbrot, Multifractality of Deutschemark/US dollar exchange rates, Cowles Foundation Discussion Paper 1166, 1997.
\bibitem{28} K. Ivanova, M. Ausloos, Eur. Phys. J. B 8 (1999) 665; Eur. Phys. J. B 12 (1999) 613 (erratum).
\bibitem{29} K. Ivanova, M. Ausloos, Physica A 265 (1999) 279.
\bibitem{30} M. Ausloos, K. Ivanova, Comp. Phys. Commun. 147 (2002) 582.
\bibitem{31} T.Di. Matteo, T. Aste, M.M. Dacorogna, Physica A 324 (2003) 183.
\bibitem{32} A. Bershadskii, Physica A 317 (2003) 591.
\bibitem{33} F. Ren, W.-X. Zhou, Europhys. Lett. 84 (2008) 68001.
\bibitem{34} J. Barunik, T. Aste, T. Di Matteo, R. P. Liu, Physica A 391(17) (2012) 4234.
\bibitem{35} W.-X. Zhou, Europhys. Lett. 88 (2) (2009) 28004.
\bibitem{36} J. de Souza, SM Duarte Queiro's, Chaos Solitons and Fractals 42(4)(2009) 2512.
\bibitem{37} W.-X Zhou, Chaos Solitons and Fractals 45 (2012) 147.
\bibitem{38} L. Zunino, B.M. Tabakd, A. Figliola, D.G. Pérez, M. Garavaglia, O.A. Rosso, Physica A, 387 (26) (2008) 6558.

%\bibitem(ft1} R. N. Mantegna, H. E. Stanley, Nature 383 (1996) 587.
\bibitem{ft22} P. Gopikrishnan, V. Plerou, Y. Liu, X. Gabaix, L.A.N. Amaral, H. E. Stanley,  Physica A 287 (2000) 362.
\bibitem{ft33} P. Gopikrishnan, V. Plerou, X. Gabaix, L. A. N. Amaral and H. E. Stanley, Physica A 299 (2001) 137.
\bibitem{ft44} K. Kiyono, Z. R. Struzik, Y. Yamamoto, Phys. Rev. Lett. 96 (2006) 068701.


%MFDFA in Finacial markets

\bibitem{mf4} K. Matia, Y. Ashkenazy, and H. E. Stanley, Europhys.Lett. 61, (2003) 422.
\bibitem{mf5} J. Kwapie\'{n}, P. O\'{s}wi\c{e}cimka, and S. Dro\.{z}d\.{z}, Physica A 350 (2005) 466.
\bibitem{mf6} P. O\'{s}wi\c{e}cimka, J. Kwapie\'{n}, and S. Dro\.{z}d\.{z}, Physica A 347 (2005) 626.
\bibitem{mf7} L. G. Moyana, J. de Souza, and S. M. D. Queiros, Physica A 371 (2006) 118.
\bibitem{mf8} J. Jiang, K. Ma, and X. Cai, Physica A 378 (2007) 399.
\bibitem{mf9} K. E. Lee and J. W. Lee, Physica A 383 (2007) 65.
\bibitem{mf10} G. Lim, S. Kim, H. Lee, K. Kim, and D.-I. Lee, Physica A 386 (2007) 259.
\bibitem{mf11} Z.-Y. Su, Y.-T.Wang, and H.-Y. Huang, J. Korean Phys.Soc. 54 (2009) 1395.
%%%%%%%%%%%%%%%%%%%%%%%%%%%%%%%%%%%%%%%%%%%%%%%%%%%%%%%%%%%%%%%%%%%%%%%%%%%%%%%%
\bibitem{oswiecimka} P. O{\'s}wi\c{e}cimka, J. Kwapie{\'n}, S. Dro{\.z}d{\.z}, A. Z. G{\'o}rski, R. Rak, Act. Phys. Pol. A 114 (2008) 3.
\bibitem{czarnecki} {\L}. Czarnecki, D. Grech, Act. Phys. Pol. A 117 (2010) 4.
\bibitem{Kantelhardt-316} J. W. Kantelhardt, S. A. Zschiegner, E. Koscielny-Bunde, S. Havlin, A. Bunde, H.E. Stanley,  Physica A 316 (2002) 87.
%%%%%%%%%%%%%%%%%%%%%%%%%%%%%%%%%%%%%%%%%%%%%%%%%%%%%%%%%%%%%%%%%%%%%%%%%%%%%%%%%%%%%%

\bibitem{58} A.Y. Schumann, J.W. Kantelhardt, Physica A 390 (2011) 2637.
\bibitem{60} Th. Lux, M. Ausloos, Market fluctuations: scaling, multi-scaling and their possible origins, in: A. Bunde, J. Kropp, H.-J. Schellnhuber (Eds.), The Science of Disaster: Scaling Laws Governing Weather, Body, Stock-Market Dynamics, Springer Verlag, Berlin (2002) 377.
\bibitem{61} S. Dro\.{z}d\.{z}, J. Kwapie\'{n}, P. O\'{s}wi\c{e}cimka, R. Rak, Europhys. Lett. 88 (2009) 60003.
\bibitem{63} D. Grech and G. Pamu{\l}a, Physica A 392 (2013) 5845.
\bibitem{64} G. Pamu{\l}a and D. Grech, Europhys. Lett. 105 (2014) 50004.
\bibitem{64a} J. Ludescher, M.I. Bogachev, J.W. Kantelhardt, A.Y. Schumann, A. Bunde, Physica A 390(13) (2011) 2480.
\bibitem{64b}  D. Gulich, L. Zunino, Physica A, 391 (16) (2012) 4100.


\bibitem{65} G. Bekaert and G. Wu, Asymmetric Volatility and Risk in Equity Markets, Review of Financial Studies, 13(1) (2000) 1.
\bibitem{66} J.-P. Bouchaud, A. Matacz, and M. Potters, Phys. Rev. Lett. 87(22) (2001) 228701.
\bibitem{67} P. T. H. Ahlgren, M. H. Jensen, and I. Simonsen, Physica A 383, (2007) 1.
\bibitem{68} {\L}. Bil, D. Grech and M. Zienowicz, PLoS ONE 12(11):e0188541(2017), https://doi.org/10.1371/journal.pone.0188541



\bibitem{seism1} L. Telesca, V. Lapenna, M. Macchiato, Physica A 354, (2005) 629.
\bibitem{seism2} L. Telesca, V. Lapenna, Tectonophysics 423 (2006) 115.
\bibitem{cosmo} M. S. Movahed, F. Ghasemi, S. Rahvar, M. R. R. Tabar, Phys. Rev. E 84 (2011) 021103.
\bibitem{biol1} P. H. Figueiredo, E. Nogueira Jr., M. A. Moret, S. Coutinho, Physica A 389 (2010) 2090.
\bibitem{biol2} S. Dutta, J. Stat. Mech. 12, (2010) 12021.
\bibitem{meteo} I. T. Pedron, J. Phys. Conf. Series 246 (2010) 012034.
\bibitem{med1} F. Liao, Y.-K. Jan, J. Rehab. Res. Develop. 48 (2011) 787.
\bibitem{med2} Y. Leung, E. Ge, Z. Yu, Ann. Assoc. Am. Geograph. 101 (2011) 1221.
\bibitem{mus1} G.R. Jafari, P. Pedram, L. Hedayatifar, J. Stat. Mech. (2007) 04012.
\bibitem{mus2} P. O\'{s}wi\c{e}cimka, J. Kwapie\'{n}, I. Celi\'{n}ska, S. Dro\.{z}d\.{z}, R. Rak, arxiv/1106.2902v1 [physics.data-an].
\bibitem{geo} F. A. Hirpa, M. Gebremichael, T. M. Over, Water Resour. Res. 46 (2010) 12529.
\bibitem{69} P. O\'{s}wi\c{e}cimka, J. Kwapie\'{n}, S. Dro\.{z}d\.{z}, Phys. Rev. E 74, (2006) 016103.



\bibitem{legendre1} J. Feder, Fractals, New York, Plenum Press (1988).
\bibitem{legendre2} H.-O. Peitgen, H. J{\"u}rgens, D. Saupe, Chaos and Fractals, $2$nd ed. Springer (2004).


\bibitem{ts1} C. Tsallis, J. Stat. Phys. 52, (1988) 479.
\bibitem{ts11} R. Rak, S.~Dro\.zd\.z, J.~Kwapie\'n, P. O\'{s}wi\c{e}cimka, Acta Physica Polonica B, 44(10) (2013) 2035.

\bibitem{ts2} C. Tsallis, R.S. Mendes, A.R. Plastino, Physica A 261  (1998) 534.
\bibitem{ts3} C. Tsallis, S.V.F. Levy, A.M.C. Souza, R. Maynard, Phys. Rev. Lett. 75  (1995) 3589.
\bibitem{ts4} D. Prato, C. Tsallis, Phys. Rev. E 60 (1999) 2398.
\bibitem{ts5} C. Tsallis, Physica A 365 (2006) 7.
\bibitem{ts6} C. Tsallis, Braz. J. Phys. 29 (1999) 1.

\bibitem{r1}  R. Rak, S.~Dro\.zd\.z, J.~Kwapie\'n, Physica A 374 (2007) 315.
\bibitem{ffm} H.A. Makse, S. Havlin, M. Schwartz, H.E. Stanley, Phys. Rev. E 53 (1996) 5445.




















\end{thebibliography}
\end{document}